\documentclass[reprint,aip,cha,twocolumn,groupedaddress,nobibnotes]{revtex4-1}
\usepackage{amsmath,amsthm,amssymb}
\usepackage[caption=false]{subfig}
\usepackage{algorithm}
\usepackage{mathtools}
\usepackage[hyperindex,breaklinks,hidelinks]{hyperref}
\usepackage{cleveref}

\DeclareMathOperator{\Ncut}{Ncut}

\DeclareMathOperator{\cut}{cut}
\begin{document}
\title{Level Set Formulation of Two-Dimensional Lagrangian Vortex Detection Methods}

\author{Alireza Hadjighasem}
\email{alirezah@ethz.ch}
\altaffiliation{Institute of Mechanical Systems, Department of Mechanical and Process Engineering, ETH Z\"{u}rich, Leonhardstrasse 21, 8092 Z\"{u}rich, Switzerland}

\author{George Haller}
\email{georgehaller@ethz.ch (Email address for correspondence)}
\altaffiliation{Institute of Mechanical Systems, Department of Mechanical and Process Engineering, ETH Z\"{u}rich, Leonhardstrasse 21, 8092 Z\"{u}rich, Switzerland}

\date{\today}

\keywords{variational level set method; Lagrangian coherent structures; nonlinear dynamical systems; vortex dynamics}

\begin{abstract}
We propose here the use of the variational level set methodology to capture Lagrangian vortex boundaries in 2D unsteady velocity fields. This method reformulates earlier approaches that seek material vortex boundaries as extremum solutions of variational problems. We demonstrate the performance of this technique for two different variational formulations built upon different notions of coherence. The first formulation uses an energy functional that penalizes the deviation of a closed material line from piecewise uniform stretching [Haller and Beron-Vera, J. Fluid Mech. 731, R4 (2013)]. The second energy function is derived for a graph-based approach to vortex boundary detection [Hadjighasem et al., Phys. Rev. E 93, 063107 (2016)]. Our level-set formulation captures an a priori unknown number of vortices simultaneously at relatively low computational cost. We illustrate the approach by identifying vortices from different coherence principles in several examples.
\end{abstract}

\maketitle

\begin{quotation}
Lagrangian Coherent structures (LCSs) such as eddies, jet streams and fronts play a vital role in various physical flows such as the atmosphere and ocean. These coherent structures are time-evolving material surfaces that split the phase-space into regions with distinct mixing and transport properties. Recent dynamical systems techniques seek such coherent structure boundaries as stationary solution of variational problems. Here, we show how these coherent structure detection methods can be reformulated such that they can be solved via the variational level set methodology.
\end{quotation}

\section{Introduction}
It has long been recognized that even temporally aperiodic flows admit emergent tracer patterns \cite{Haller15}. Generally referred to as Lagrangian coherent structures (or LCS), these patterns are often vortex-type (or elliptic) spatial features that remain recognizable over times exceeding typical time scales in the flow. Such elliptic LCSs arise in a number of natural phenomena, ranging from Jupiter's mysterious Great Red Spot to mesoscale eddies that populate nearly all parts of the global ocean.

Lagrangian (i.e., trajectory-based) vortex detection approaches can roughly be divided into three categories: geometric, set-based and diagnostic methods. Geometric methods  identify vortex boundaries as either outermost non-filamenting, closed material surfaces \cite{Haller13,Haller15,Hadjighasem16} or as outermost, closed material surfaces of equal material rotation \cite{Haller16,Farazmand15}. In contrast, set-based approaches aim to detect the interiors of coherent flow regions, as opposed to the boundaries encompassing these regions. Examples include probabilistic methods for detecting almost-invariant and finite-time coherent sets \cite{Froyland14}; ergodicity-based methods for time-periodic flows \cite{Budisic12_1}; braid-theoretical methods for flows with recurrent trajectories \cite{Allshouse12}; and trajectory clustering approaches \cite{Hadjighasem16_2,Froyland15} for aperiodic flows. Finally, diagnostic approaches propose Lagrangian scalar fields whose distribution is expected to reflect coherent features of the flow \cite{Rypina11,Mezic10}. Unlike the first two categories, diagnostic methods offer no well-defined boundaries for their vortical features. The level-set approach developed here falls in the first category, focusing on the precise identification of Lagrangian vortex boundaries from variational principles.

Since its introduction, the level set method has widely been applied within different fields of science. These include optimal-time path planning \cite{Lolla14}; image processing  \cite{Vese02,Zhao01,Rudin92}; two phase flow simulation \cite{Sussman94}; fluid-interface problems \cite{Sethian03}; finite-time Lyapunov exponent (FTLE) calculation \cite{Leung11}; limit cycle detection \cite{You15}; and ergodic partitioning of continuous
dynamical systems \cite{You14}. In some of the these applications, level set functions are used to partition a domain into qualitatively different regions. The dynamic interfaces separating those regions are marked by zero sets of a level set function. The interface motion is often determined by partial differential equations derived from physical principles, e.g., the propagation of a flame front in a combusting gas \cite{Fedkiw99}. In other cases, however,  the evolution equation of the dynamic interface is derived from the problem of minimizing a certain energy functional defined on level sets. These types of level set methods are known as \emph{variational level set methods} \cite{Tsai03}. 

In this paper, we apply the variational level-set methodology to partition fluid domains into coherent and incoherent regions. Specifically, we present two variational formulations that force the level sets to evolve toward vortex boundaries in the flow. Our first formulation seeks boundaries of coherent Lagrangian vortices as closed material lines that exhibit nearly uniform stretching. This formulation builds on the geodesic LCS principle \cite{Haller13} that identifies vortex boundaries as outermost members of uniformly stretching closed material curve families. Our second formulation seeks coherent vortices as patches of Lagrangian particles that evolve most tightly under the action of the fluid flow in space-time \cite{Hadjighasem16_2}. Both approaches lead to energy functionals whose minima describe coherent vortex boundary curves. Using calculus of variations, we then derive a gradient flow that minimizes each energy functional over a space of level-set functions. This gradient flow in turn drives the motion of an arbitrary closed initial curve, defined implicitly as a zero level set of a function, toward vortex boundaries.

The variational level set methodology proposed here has three main advantages. First, it  captures an a priori unknown number of vortices for automated vortex census and tracking (see also Refs.~\onlinecite{Hadjighasem16_2,Haller16,Karrasch15}). Second, the method carries out the computation over a limited number of pixels, hence its computational cost does not scale up drastically with the resolution of the computational domain. This feature renders the level set method a viable approach for tackling high-resolution data sets. Finally, the methodology can be adapted to any other variational coherent structure detection principle. Here, we specifically demonstrate a stretching- and a graph-based energy functional, but applications to other coherence principles are equally possible.

The rest of this paper is organized as follows. \Cref{section:background} briefly reviews the necessary background on the standard level set method. In \cref{section:LevelsetFormulation}, we develop two new formulations for identifying coherent Lagrangian vortices within the variational level set framework. In \cref{appendix:Implementation}, we discuss the numerical aspects of our proposed method along with a detailed numerical implementation. Finally, we illustrate our results on several examples, ranging from analytic velocity fields to time-dependent two-dimensional observational data in \cref{section:results}.

\section{Background}\label{section:background}
\subsection{Implicit boundary representation}
\begin{figure*}
\subfloat[]{\includegraphics[height=0.35\textwidth]{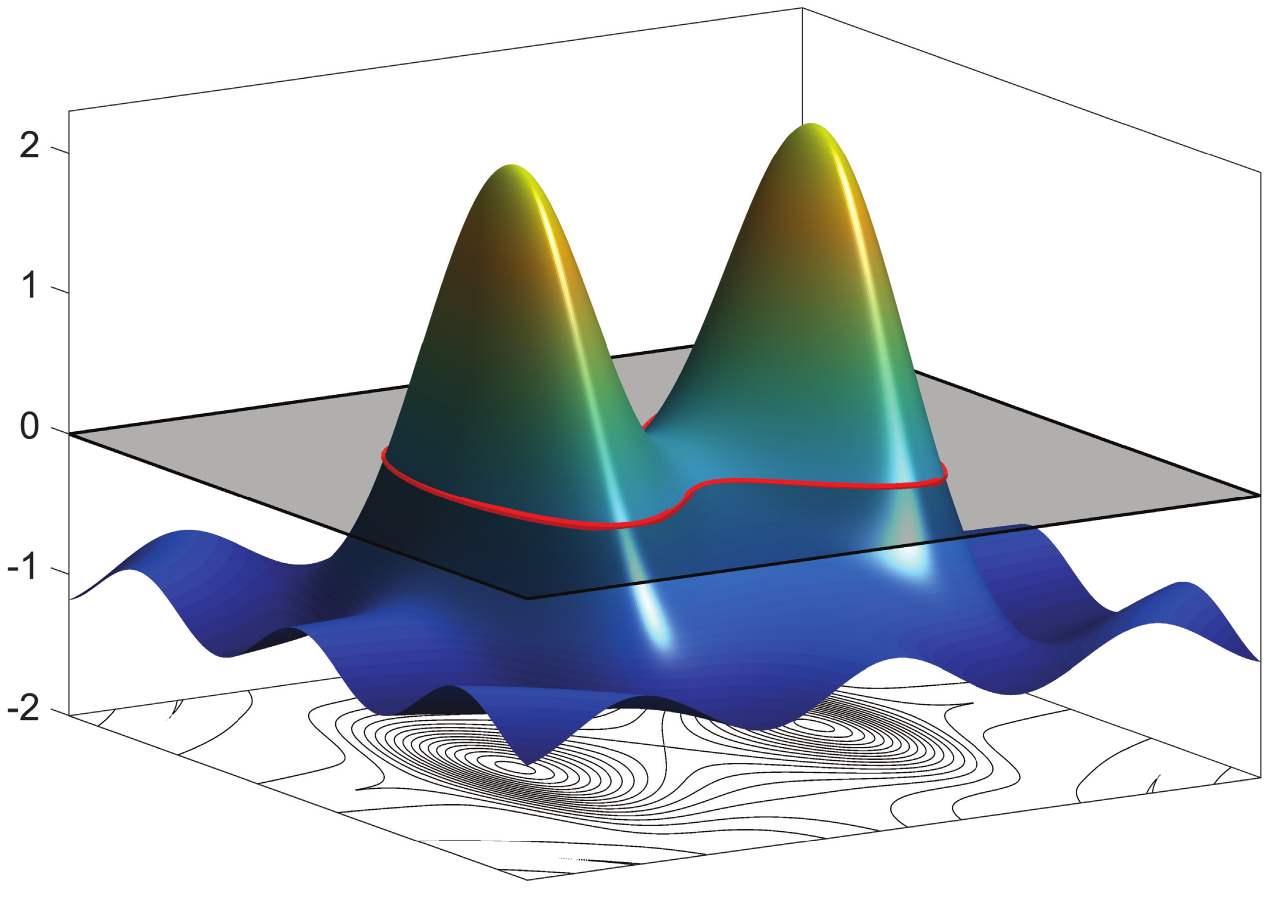}}\quad
\subfloat[]{\includegraphics[height=0.3\textwidth]{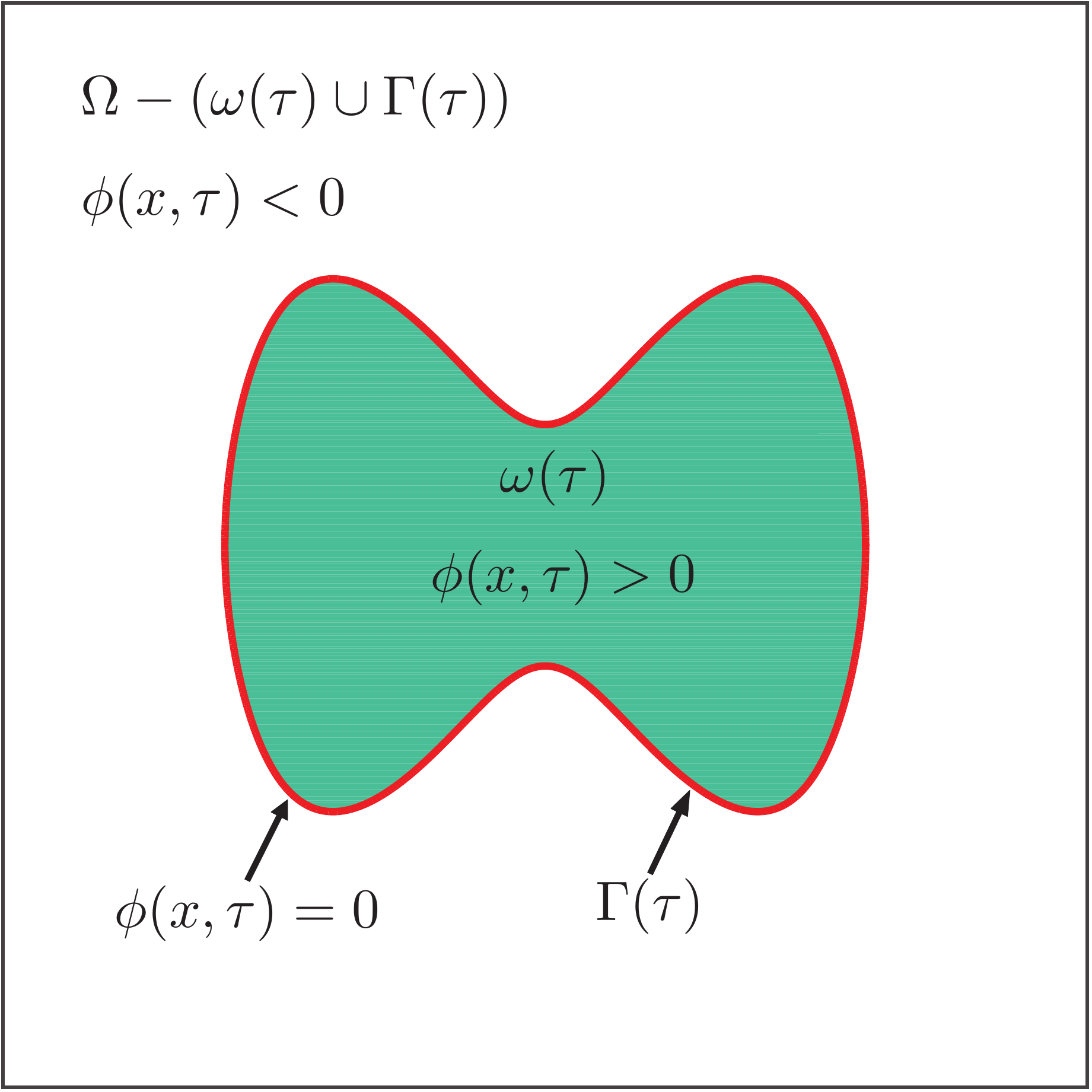}}
\caption{(a) Level set function and its zero level contour (red). (b) A curve $\Gamma$, implicitly represented by the zero level set of the function $\phi$, is the boundary between the regions $\lbrace(x,y) : \phi(x,y) > 0\rbrace$ and $\lbrace (x,y) : \phi(x, y) < 0\rbrace$.}
\label{fig:LevelSetTopology}
\end{figure*}

We begin by reviewing the standard level set method, as devised by Osher \& Sethian \cite{Osher88}. Consider a closed moving interface as a curve $\Gamma(\tau)$ in $\mathbb{R}^2$, with $\tau$ denoting the time of evolution. Let $\omega(\tau)$ be the open region that $\Gamma(\tau)$ encloses in the domain $\Omega$ (see \cref{fig:LevelSetTopology}). The main idea of the level set methodology is to embed $\Gamma(\tau)$ as the zero-level set of a higher-dimensional function $\phi(\cdot,\tau):\Omega\rightarrow\mathbb{R}$, called \emph{the level set function}, which is assumed Lipschitz continuous and satisfies the following conditions: 
\begin{equation*}
\begin{aligned}
\phi(x,\tau) > 0 & \quad \text{for $x\in  \omega(\tau)$},\\
\phi(x,\tau) < 0 & \quad \text{for $x \in \Omega-\left(\omega(\tau) \cup\Gamma(\tau)\right)$},\\
\phi(x,\tau) = 0 & \quad \text{for $x \in \Gamma(\tau)$}.
\end{aligned}
\end{equation*}
Conversely, if we know $\phi(x,\tau)$, we may locate the interface by finding the zero level set of $\Gamma(\tau)=\lbrace x:\phi(x,\tau)=0\rbrace$. Evolving the interface $\Gamma(\tau)$ in $\tau$ is equivalent to updating $\phi(x,\tau)$.

A typical example of a level set function is given by the \emph{Signed Distance Function} (SDF) measured from a curve. The SDF computed for $\Gamma(\tau)$ gives the distance of a given point $x$ from the interface $\Gamma(\tau)$, with the sign determined by whether $x$ is inside or outside $\omega(\tau)$. The SDF has positive values inside $\Gamma(\tau)$, decreases to zero as $x$ approaches $\Gamma(\tau)$, and takes negative values outside of $\Gamma(\tau)$. Signed distance functions share all the properties of implicit functions, such as supporting Boolean operations (union, intersection, and difference), in addition to the identity $\left|\nabla\phi\right| = 1$ (cf. Ref.~\onlinecite{Osher06}).

\subsection{Front evolution and level set theory}\label{section:LevelsetTheory}
Given an interface $\Gamma(\tau)$, our goal is to produce an equation for evolving $\phi(x,\tau)$, as the embedding of $\Gamma(\tau)$, through space and time such that the interface  $\Gamma(\tau)$ advances toward the vortex boundaries. The variational level set approach obtains the equations governing the evolution of $\Gamma(\tau)$ by minimizing a certain energy functional $E$ defined on the level set function $\phi(x,\tau)$. The energy functional $E$ can depend on the intrinsic geometric properties of the interface (e.g, curvature) or on extrinsic quantities (e.g., velocity of the fluid flow). The spatio-temporal partial differential equation describing the evolution of the level set function is given by
\begin{equation}
\frac{\partial\phi}{\partial\tau} = -\frac{\partial E}{\partial\phi}. 
\label{eq:LevelSetEvolution}
\end{equation}
\Cref{eq:LevelSetEvolution} is a \emph{gradient flow} \cite{Evans10} that minimizes the functional $E$ and simultaneously governs the evolution of the interface $\Gamma(\tau)$. There are several advantages associated with this perspective:
\begin{enumerate}
\item Although $\phi(x,\tau)$ remains a smooth function, the level surface $\phi(x,\tau) = 0$ corresponding to the propagating interface may develop sharp corners, break apart, or merge. No elaborate numerical mechanism is required to handle such topological changes.
\item The level set function always remains a function on a fixed grid, which allows for efficient numerical schemes.
\item Intrinsic geometric properties of the interface $\Gamma(\tau)$ are obtained directly from the level set function $\phi$. For instance, the outward unit normal vector to $\Gamma(\tau)$ is given by $n = \frac{\nabla\phi}{\left|\nabla\phi\right|}$, and the mean curvature of each level set is $\kappa = \nabla\cdot\frac{\nabla\phi}{\left|\nabla\phi\right|}$. Other geometric quantities, such as the arclength $|\Gamma|$ and the enclosed area $|\omega|$ of $\omega$, can be expressed respectively as \cite{Chang96,Osher06}:
\begin{equation}
\begin{split}
|\Gamma| &= \int_\Omega \delta(\phi(x,\tau))\left|\nabla\phi(x,\tau)\right|\;dx, \\
|\omega| &= \int_{\Omega} H(\phi(x,\tau))\;dx, 
\end{split}
\end{equation}
where $H(\phi)$ is the Heaviside function, and $\delta(\phi)$ is its derivative, the Dirac delta function. 
\end{enumerate}
We shall omit the dependence of $\phi$ on the spatial variable $x$ and the evolution time $\tau$ for notational simplicity.

\section{Variational level-set-based vortex detection}\label{section:LevelsetFormulation}
In the previous section, we discussed how to represent a curve implicitly and advect it with a gradient flow using the level set method. We have not yet discussed, however, how the energy function can be constructed to ensure that an arbitrary closed curve moves towards vortex boundaries. As we shall see below, such an energy functional should have local minima that mark the desired vortex boundaries. We work out the derivations of two energy functionals for detecting vortex boundaries. Our first derivation relies on the uniform stretching properties of Lagrangian vortex boundaries. Our second functional characterizes vortex boundaries based on the sustained proximity of Lagrangian particles in the spatio-temporal domain these boundaries enclose.

\subsection{Stretching-based formulation}\label{section:StretchingFormulation}
We start with an unsteady velocity field
\begin{equation}
\dot{x}=v(x,t),\quad x\in U\subset\mathbb{R}^{2},\quad t\in[t_{0},t_{1}],
\label{eq:dynsys}
\end{equation}
which defines a two-dimensional flow over the finite time interval $[t_{0},t_{1}]$ in the spatial domain $U$. The flow map $F_{t_{0}}^{t_{1}}(x_{0}):x_{0}\mapsto x_{t_{1}}$ of \eqref{eq:dynsys} then maps the initial condition $x_{0}$ at time $t_{0}$ to its evolved position $x_{t_{1}}$ at time $t_{1}$. The right Cauchy--Green (CG) strain tensor associated with \eqref{eq:dynsys} is defined as \begin{equation} C_{t_{0}}^{t_{1}}(x_{0})=\left({\nabla F_{t_{0}}^{t_{1}}}\right)^{\intercal}\nabla F_{t_{0}}^{t_{1}}, \end{equation} where $\nabla F_{t_{0}}^{t_{1}}$ denotes the gradient of the flow map, and the symbol $\intercal$ indicates matrix transposition. We shall suppress the dependence of CG on $t_{0}$ and $t_{1}$ for notational simplicity. 

We seek a Lagrangian vortex boundary as an exceptional closed material
line $\Gamma$ around which $\mathcal{O\left(\epsilon\right)}$-thick coherent belts
show minimal variation in the length-averaged Lagrangian strain over the time interval $[t_{0},t_{1}]$. This view is motivated by Ref.~\onlinecite{Haller13}, where the authors seek a perfectly coherent boundary as a material line exhibiting no leading order variation in material strain across the $\mathcal{O\left(\epsilon\right)}$-thick coherent belts. Solutions to this variational problem turn out to be closed material lines that are infinitesimally uniformly stretching, i.e., all their subsets stretch by the same amount between the times $t_{0}$ and $t_{1}$. Compared to Ref.~\onlinecite{Haller13}, we do not explicitly enforce such uniform stretching, but require the vortex boundary to have as little nonuniformity in its stretching as possible. In \cref{section:Jupiter}, we will apply both the original coherence principle \cite{Haller13} and its present relaxed version to identify the boundary of the Great Red Spot (GRS) in Jupiter's atmosphere.

To express our stretching-based energy functional mathematically, we select a parametrization $r(s)$ with $s\in[0,\sigma]$ for the closed $\Gamma$. We let $l_{t_{0}}(s)$ denote the length of a tangent vector $r'(s)$ at initial time $t_{0}$, and $l_{t_{1}}(s)$ denote the length of the corresponding tangent vector at final time $t$. These two tangent lengths can be calculated as \cite{Truesdell04}:
\begin{equation}
\begin{aligned}
l_{t_{0}} &= \sqrt{\left<r^{\prime}\left(s\right),r^{\prime}\left(s\right)\right>},\\
l_{t_{1}} &= \sqrt{\left<r^{\prime}\left(s\right),C\left(r\left(s\right)\right)r^{\prime}\left(s\right)\right>}.
\label{eq:length}
\end{aligned}
\end{equation}
The quadratic variation of tangential strain along $\Gamma$ is then given by
\begin{equation}
E(\Gamma,c_{0}) = \int_{\Gamma} \left(\frac{l_{t_{1}}(s)}{l_{t_{0}}(s)}-c_0 \right)^{2}ds,
\label{eq:functional_s}
\end{equation}
where $c_{0}$ is an unknown constant to be determined. Expressing the interface $\Gamma$ implicitly as the zero level set of a function $\phi$, we obtain
\begin{equation}
E(\phi,c_{0}) = \int_{\Omega} f(\phi,\nabla\phi,c_{0})\;dx,
\label{eq:functional_phi}
\end{equation}
where
\begin{equation}
f(\phi,\nabla\phi,c_{0}) = \left(\frac{1}{\left|\nabla\phi\right|}\sqrt{\left<\nabla\phi,\tilde{C}\nabla\phi\right>}-c_0 \right)^{2}\delta(\phi),
\end{equation}
and $\tilde{C} = R_{\pi/2}^{\intercal} C R_{\pi/2}$, with $R_{\pi/2}$ referring to a counter-clockwise rotation by $\pi/2$.

Equation \eqref{eq:functional_phi} is a multivariable functional that can be minimized via the \emph{alternative optimization} \cite{Csiszar84} procedure as follows. First, we fix $\phi$ to optimize for $c_0$ and then fix $c_0$ for optimizing over $\phi$. When $\phi$ is fixed, we obtain the optimum
\begin{equation}
c_0 = \frac{1}{\sigma}\int_{\Omega} \frac{1}{\left|\nabla\phi\right|}\sqrt{\left<\nabla\phi,\tilde{C}\nabla\phi\right>} \delta(\phi)\;dx\equiv \frac{1}{\sigma}\int_{\Gamma} \frac{l_{t_{1}}(s)}{l_{t_{0}}(s)}ds,
\label{eq:average_stretching}
\end{equation}
which is just the average relative stretching along the curve $\Gamma$. Keeping this $c_{0}$ fixed and formally optimizing the energy with respect to $\phi$, we obtain the Euler-Lagrange equation
\begin{widetext}
\begin{equation}
\begin{split}
\frac{\partial E}{\partial\phi} &= \frac{\partial f}{\partial\phi}-\nabla\cdot \frac{\partial f}{\partial\nabla\phi} = 0,\\
\frac{\partial E}{\partial\phi} &= -2\delta(\phi)\nabla\cdot\left[\left(c_{0}\frac{\sqrt{\left<\frac{\nabla\phi}{\left|\nabla\phi\right|},\frac{\tilde{C}\nabla\phi}{\left|\nabla\phi\right|}\right>}}{\left|\nabla\phi\right|}-\frac{\left<\frac{\nabla\phi}{\left|\nabla\phi\right|},\frac{\tilde{C}\nabla\phi}{\left|\nabla\phi\right|}\right>}{\left|\nabla\phi\right|}\right)\frac{\nabla\phi}{\left|\nabla\phi\right|}+\left(1-\frac{c_{0}}{\sqrt{\left<\frac{\nabla\phi}{\left|\nabla\phi\right|},\frac{\tilde{C}\nabla\phi}{\left|\nabla\phi\right|}\right>}}\right)\frac{\tilde{C}\nabla\phi}{\left|\nabla\phi\right|^{2}}\right],
\label{eq:functional_derivative}
\end{split}
\end{equation}
\end{widetext}
with the Neumann boundary conditions \cite{Evans10} imposed on the domain boundaries.

To find the minimum of $E$ with respect to $\phi$ numerically, we parameterize the descent direction by an artificial time $\tau\geq 0$, and solve the gradient descent \cref{eq:LevelSetEvolution}. The total energy \eqref{eq:functional_phi} is then minimized by iterating the contour evolution \eqref{eq:LevelSetEvolution} in alternation with the update \eqref{eq:average_stretching} of the average stretching parameter. 

\subsection{Graph-based formulation}\label{section:GraphFormulation}

\begin{figure*}
\subfloat[]{\includegraphics[height=0.3\textwidth]{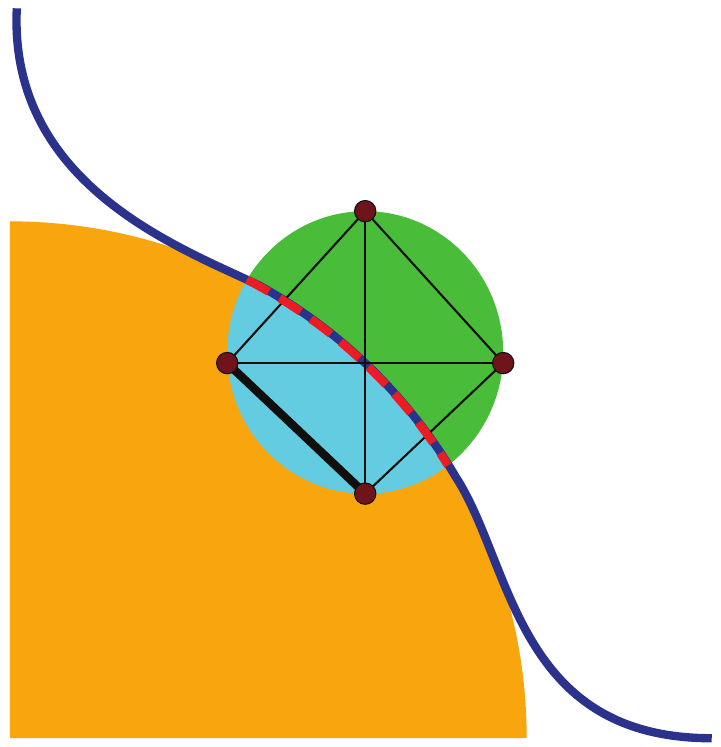}}\quad
\subfloat[]{\includegraphics[height=0.3\textwidth]{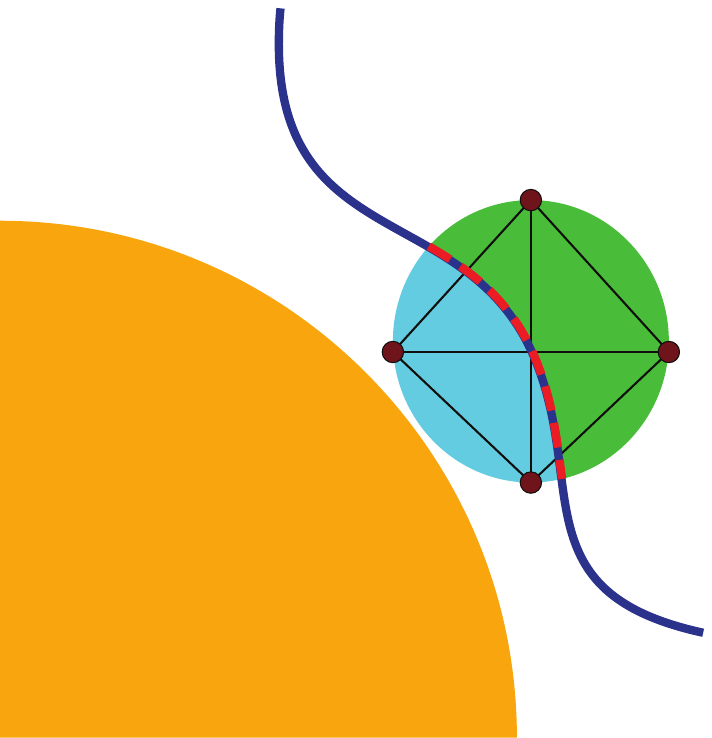}}\quad
\subfloat[]{\includegraphics[height=0.3\textwidth]{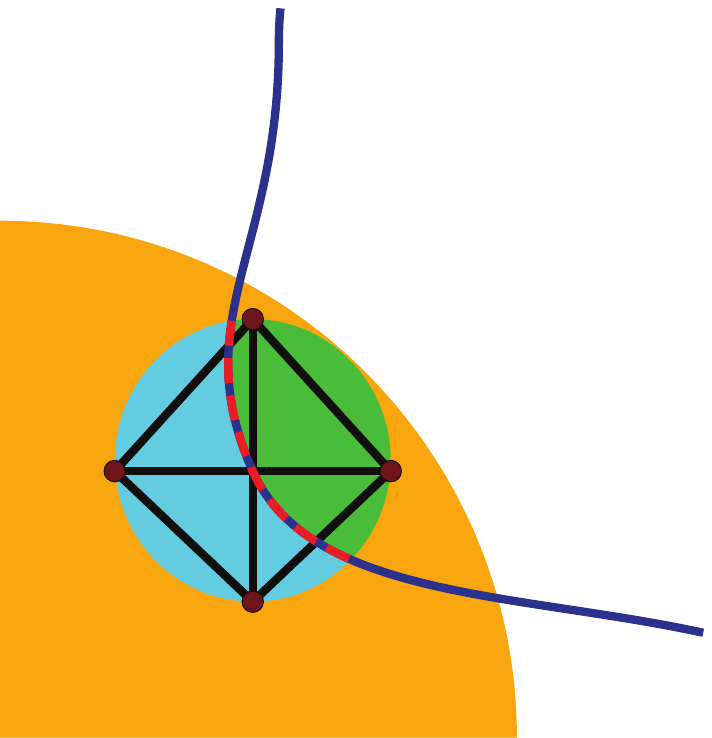}}
\caption{The optimality of the Normalized Cut value for three different scenarios. (a) localized graph is centered in the vicinity of a vortex boundary (orange). (b) localized graph is centered far away from the the vortex boundary. (c) localized graph is centered inside the vortex. The evolving zero level set is illustrated in dark blue.}
\label{fig:LocalizedGraph}
\end{figure*}

In this section, we describe an alternative approach to vortex identification that relies on spectral graph theory \cite{Chung97} and a localized level set model \cite{Lankton08}.  Within this framework, the contour moves based on the localized energies obtained directly from nearby particle trajectories. To compute these local energies, we form small regions around each point along the evolving curve such that each region is split into a local interior and a local exterior by the curve (see \cref{fig:LocalizedGraph}). We then obtain the level-set evolution equation by optimizing a functional that incorporates these local energies. Below we describe this approach in more detail using related concepts from Refs.~\onlinecite{Hadjighasem16_2,Lankton08}.

We start by defining a second spatial variable $y$ that also labels points in $\Omega$. We then define a mask function $B(x,y)$, that acts as an indicator function for points $x$ and $y$ within a distance $R$ \cite{Lankton08}:
\begin{equation*}
    B(x,y)= 
\begin{cases}
    1,& \text{if }  \lVert x-y \rVert < R\\
    0,              & \text{otherwise}.
\end{cases}
\end{equation*}
The function $B(x,y)$ is, therefore, equal to $1$ when the point $y$ is within a ball of radius $R$ centered at $x$, and is equal to $0$ otherwise.\\
The associated localized energy along an evolving curve $\Gamma$ is then given by
\begin{equation}
E(\phi) = \int_{\Omega_{x}} \delta \left(\phi(x)\right) \int_{\Omega_{y}} B(x,y)\cdot \mathcal{F}(\phi(y))\;dy\;dx,
\label{eq:Functional_graph}
\end{equation}
where $\mathcal{F}$ is a function designed to detect the presence of vortex boundaries within a $B(x,y)$ neighborhood of a point on the evolving curve $\Gamma$. We then optimize the energy functional \eqref{eq:Functional_graph} by taking its first variation with respect to $\phi$ as follows (see Ref.~\onlinecite{Lankton08} for more details)
\begin{equation}
\frac{\partial E}{\partial\phi} = \delta\left(\phi(x)\right)\int_{\Omega_{y}} B(x,y)\cdot \nabla_{\phi(y)}\mathcal{F}(\phi(y))\;dy.
\label{eq:graph_fun_der}
\end{equation}
Here, we propose $\mathcal{F}$ to be the \emph{normalized cut} or \emph{Ncut} \cite{Shi00} value obtained from bi-partitioning of a similarity graph $G$ built locally in a $B(x,y)$ neighborhood of each point on the evolving curve $\Gamma$ (see \cref{fig:LocalizedGraph}). To construct the similarity graph $G$, we follow the procedure specified in Ref.~\onlinecite{Hadjighasem16_2}.

In short, we define the similarity graph $G=(V,E,W)$ through the set of its nodes $V=\left\{ v_{1},...,v_{n}\right\}$, the set of its edges $E\subseteq V\times V$ between nodes, and a similarity matrix $W\in\mathbb{R}^{n\times n}$ which associates weight $w_{ij}$ to the edge $e_{ij}$ between the nodes $v_{i}$ and $v_{j}$. In our context, we interpret the graph nodes $V$ as a set of Lagrangian particles released within $B(x,y)$, and the associated similarity weights $w_{ij}$ as the inverse of the average Euclidean distance between particle trajectories. We compute this average Euclidean distance using the \emph{dynamic distance metric} \cite{Hadjighasem16_2}.

The Ncut graph-clustering algorithm seeks to partition the nodes $V$ into a set $A$ and its complement $\bar{A}$, such that both of the following hold:
\begin{description}
\item[Within-cluster similarity] Nodes in the same cluster are similar to each other, i.e.,  particles in a coherent structure have mutually short dynamical distances.
\item[Between-cluster dissimilarity] Nodes in a cluster are dissimilar from those in the complementary cluster, i.e., particles in a coherent structure are expected to have long dynamical distances from the rest of the particles.
\end{description}
The normalized cut that directly implements the above (dis)similarity conditions can be formulated mathematically as
\begin{equation}
\begin{aligned}
\Ncut(A,\bar{A})&=\frac{\cut(A,\bar{A})}{\cut(A,V)}+\frac{\cut(A,\bar{A})}{\cut(\bar{A},V)},\\
\cut(A,\bar{A})&=\sum_{u\in A, v\in \bar{A}}w(u,v).
\end{aligned}
\label{eq:Ncut}
\end{equation}
Additional minor details for implementing the graph cut algorithm such as sampling trajectories over discrete times and sparsifying the similarity graph are discussed in detail in Ref.~\onlinecite{Hadjighasem16_2}.

With this definition, we now argue that the value of Ncut is locally minimum when the localized graph is centered in the vicinity of a vortex boundary. To clarify this further, we discuss the optimality of Ncut value for three plausible scenarios: localized graph is centered in the vicinity of a vortex boundary, inside  the mixing region and inside a vortex (see \cref{fig:LocalizedGraph}). In the first scenario, the value of Ncut is small since the graph can be split into a cluster $A$ and its complement $\bar{A}$ such that the edges between $A$ and $\bar{A}$ have low weights and the edges within $A$ have high weights. In contrast, the Ncut value will be large inside the mixing region since the edges within $A$ will have low weights. We also expect that the value of Ncut will be large inside the vortex as well because all nodes are strongly connected. This means that the evolving level set function $\phi$ becomes trapped at vortex boundaries, given that the energy functional \eqref{eq:Functional_graph} is locally minimal.

\section{Numerical results}\label{section:results}
We now summarize our algorithms for detecting coherent Lagrangian vortices using stretching- and graph-based formulations in the tables entitled \Cref{alg:algorithm1} and \Cref{alg:algorithm2} below.

\begin{algorithm}[H]
\caption{Stretching-Based Level Set Method}
\label{alg:algorithm1}
\begin{enumerate}
\item \textit{Initialization}:
\begin{enumerate}
\item Generate a sufficiently large closed curve and initialize the level set function $\phi$ as a signed distance function $\phi_{0}$ measured from this curve.
\item Construct the active set $L_{0}$ and populate the neighbor layers $L_{i}$ by determining the distance of a neighborhood point from the nearest active point (see \cref{appendix:SFM}).  
\end{enumerate}
\item \textit{Update the zero level set}: 
\begin{enumerate}
\item Compute the gradient flow using \eqref{eq:functional_derivative} for the active set $L_{0}$.
\item Evolve the active set with \eqref{eq:LevelSetEvolution} to time $\tau_{k+1} = \tau_{k}+\Delta\tau$ such that $\Delta\tau$ satisfies the CFL condition (cf. \cref{appendix:Implementation}).
\end{enumerate}
\item \textit{Update the sparse band}: Update the level set location and the corresponding neighboring layers $L_{i}$.
\item \textit{Convergence}: Check whether the iterations have converged. If yes, stop; otherwise go to step 2. 
\end{enumerate}
\end{algorithm}

\begin{algorithm}[H]
\caption{Graph-Based Level Set Method}
\label{alg:algorithm2}
\begin{enumerate}
\item \textit{Initialization}:
\begin{enumerate}
\item Generate a sufficiently large closed curve and initialize the level set function $\phi$ as an SDF.
\item Construct the active set $L_{0}$ and populate the neighbor layers $L_{i}$.
\end{enumerate}
\item \textit{Update the zero level set}: 
\begin{enumerate}
\item Construct a localized graphs for the active set 
\item Calculate the Ncut for each localized graph $G$ such that the graph  will be partitioned into a local interior and a local exterior by the curve.
\item Compute the gradient flow using \eqref{eq:graph_fun_der} for the active set $L_{0}$.
\item Evolve the active set to time $\tau_{k+1} = \tau_{k}+\Delta\tau$ such that $\Delta\tau$ satisfies the CFL condition, and the total energy decreases.
\end{enumerate}
\item \textit{Update the sparse band}: Update the level set location and the corresponding neighboring layers $L_{i}$.
\item \textit{Convergence}: Check whether the iterations have converged. If yes, stop; otherwise go to step 2. 
\end{enumerate}
\end{algorithm}
The computational cost of our implementation is primarily due to step 2, i.e. the construction of the Cauchy--Green strain tensor or the localized graph for the active set. This accounts for about $75-95\%$ of the total execution time, depending on the  perimeter length of the zero level set and the resolution of the grid.

We demonstrate the implementation of \Cref{alg:algorithm1,alg:algorithm2} on three examples to detect coherent Lagrangian vortices. In the first example, we consider a periodically forced pendulum for which we can explicitly confirm our results using an appropriately defined Poincar\'{e} map. Our second example, Jupiter's unsteady wind-velocity field has a higher-level temporal complexity. In this example, we use a time-resolved velocity field reconstructed from an enhanced video footage of Jupiter, capturing Jupiter's Great Red Spot (GRS) \cite{Hadjighasem16}. In the third example, we detect coherent Lagrangian vortices in a quasigeostrophic ocean surface flow derived from satellite-based sea-surface height observations \cite{Fu10}.

To implement \Cref{alg:algorithm1,alg:algorithm2} in the forthcoming
examples, we use a variable-order Adams-Bashforth-Moulton solver (ODE113
in MATLAB) to advect fluid particles with the differential equation \eqref{eq:dynsys}. The absolute and relative tolerances of the ODE solver are chosen as $10^{-6}$. In \cref{section:Ocean,section:Jupiter},
we obtain the velocity field at any given point by interpolating the
velocity data set using bilinear interpolation.

To evolve the level set function, we use an explicit time-marching scheme governed by the CFL condition (see \cref{appendix:Implementation}). We choose the corresponding CFL number $\mu=0.5$ and the regularization parameter $\varepsilon = 10^{-4}$, unless stated otherwise. Moreover, we initiate the level set evolution with a large enough closed curve that is expected to encircle all coherent Lagrangian vortices. We then evolve the level set function inward so as to capture the coherent vortices individually.

\subsection{Periodically forced pendulum}\label{section:Pendulum}
\begin{figure*}
\subfloat[\label{fig:pendulum1}]{\includegraphics[height=0.33\textwidth]{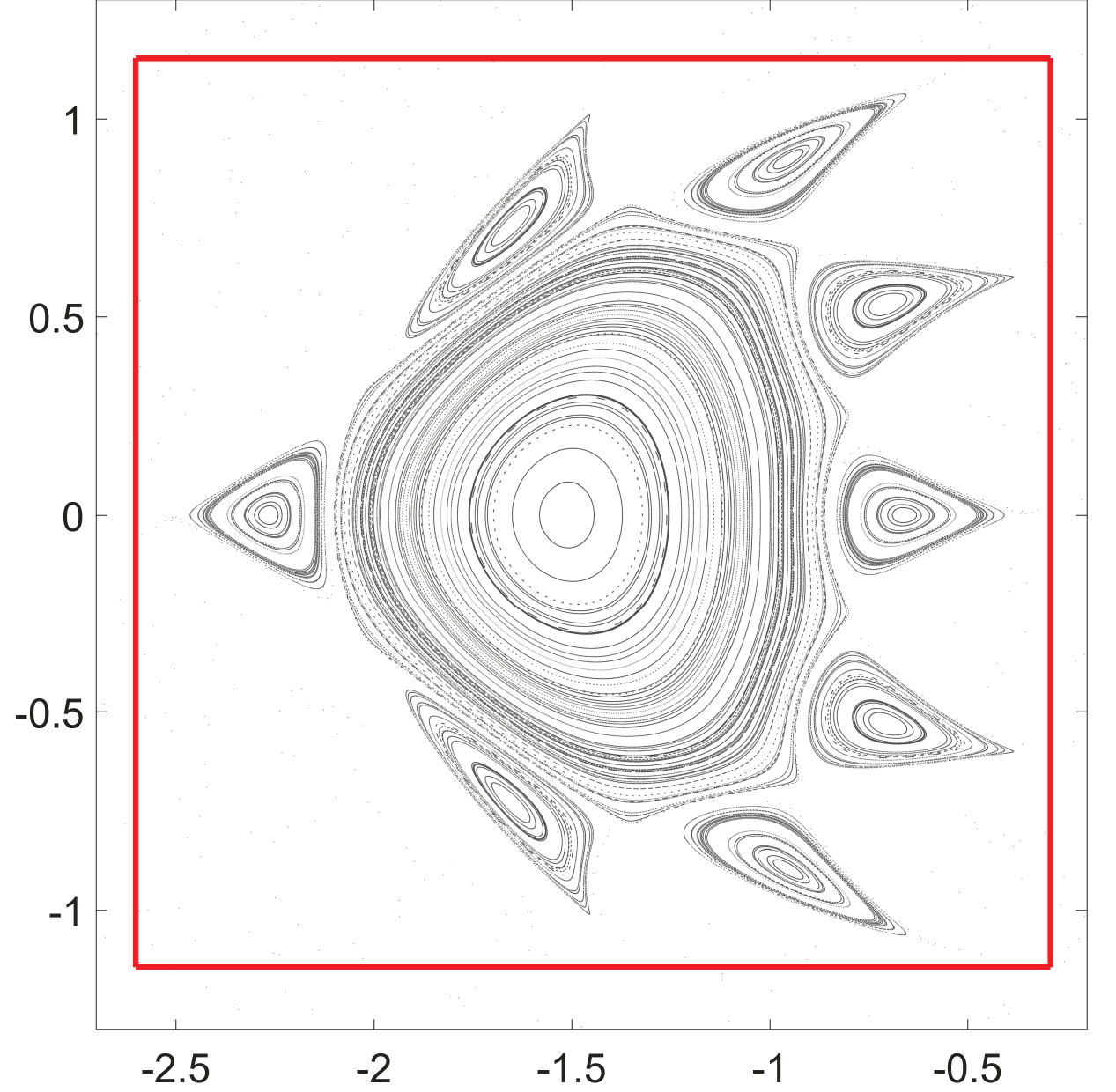}}
\subfloat[\label{fig:pendulum2}]{\includegraphics[height=0.33\textwidth]{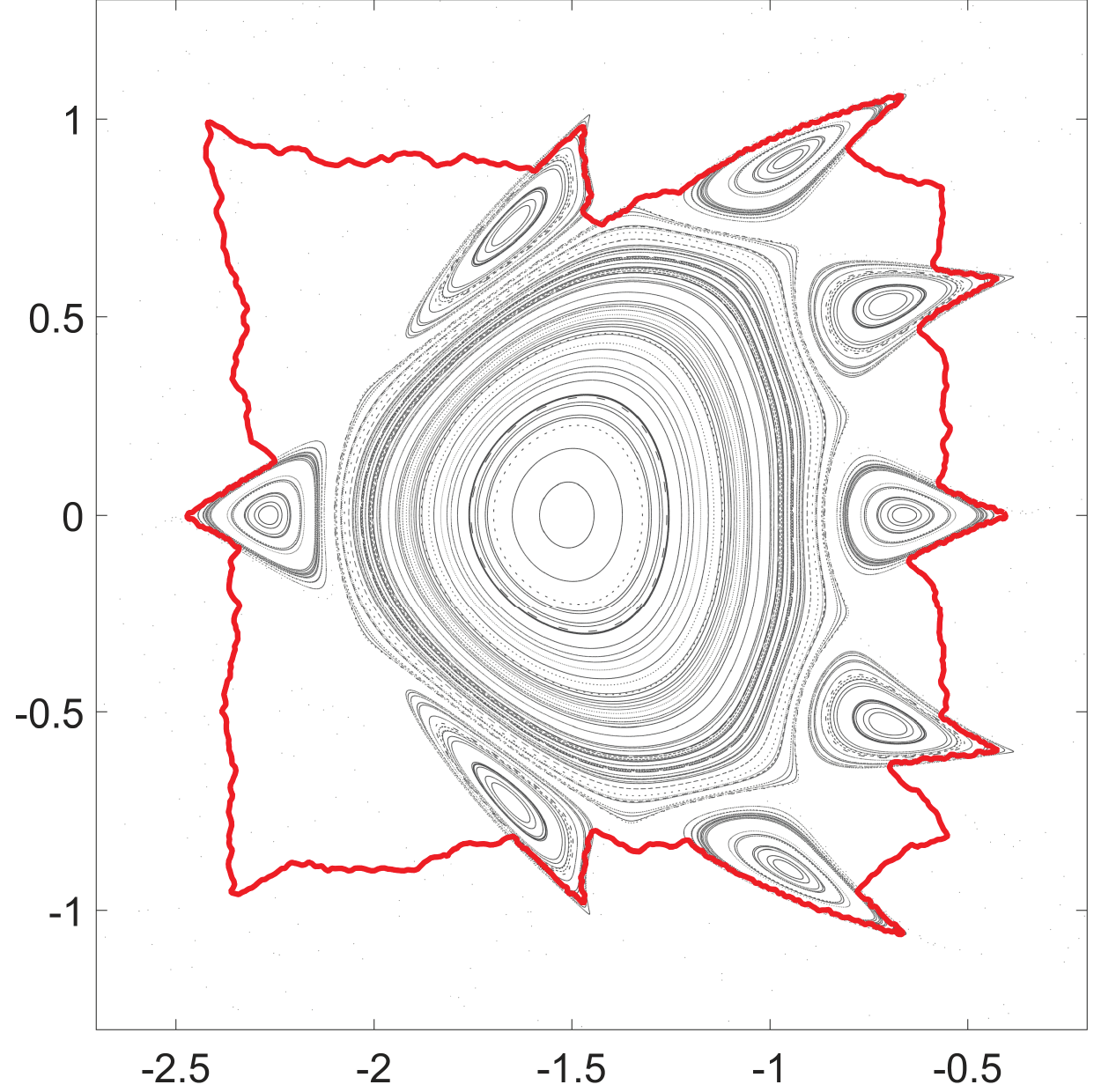}}
\subfloat[\label{fig:pendulum3}]{\includegraphics[height=0.33\textwidth]{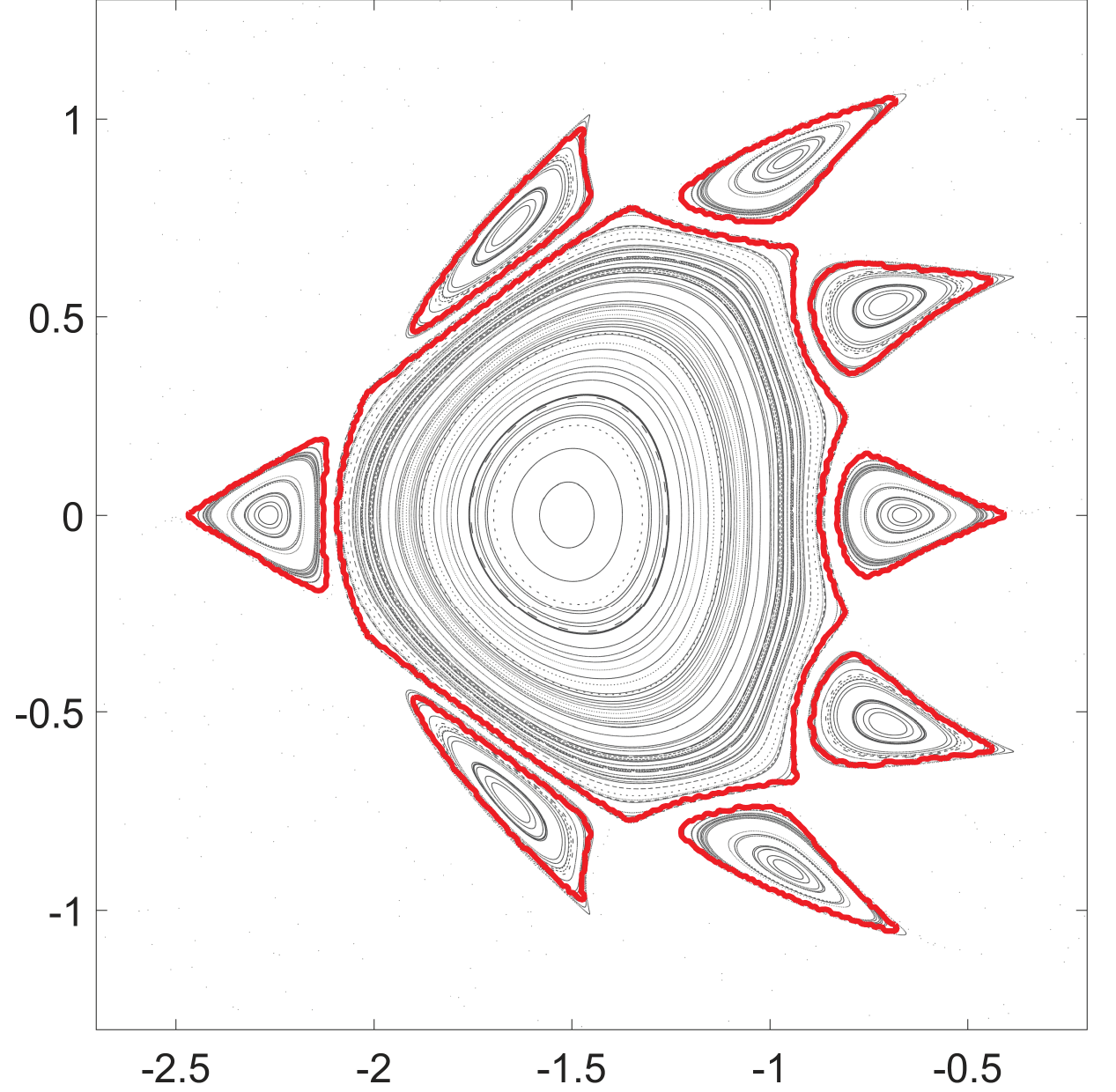}}
\caption{(a-c) Evolution of the zero level set toward the boundary of KAM regions for the periodically forced pendulum. Shown in the background is a Poincar\'{e} map constructed for 800 iterations. (Multimedia view)}
\label{fig:PendulumLevelsetEvolution}
\end{figure*}

Consider the periodically forced pendulum
\begin{equation*}
\begin{aligned}\dot{x}_{1} & =x_{2}\\
\dot{x}_{2} & = -\sin x_{1}+\epsilon\cos t.
\end{aligned}
\end{equation*}

For $\epsilon=0$, the system is integrable, and has chains of alternating elliptic and hyperbolic fixed points, with periodic orbits encircling the elliptic fixed points, and heteroclinic orbits connecting adjacent hyperbolic fixed points. These orbits form invariant sets on the Poincar\'{e} map $\mathcal{P}\coloneqq F_{0}^{2\pi}$.

For $0<\epsilon\ll 1$, however, the closed invariant sets for $\mathcal{P}$ generally break up. We set the perturbation strength to $\epsilon=0.4$ and reveal the surviving KAM regions by constructing the Poincar\'{e} map $\mathcal{P}$ for 800 iterations. A similar parameter setting was also studied in Ref.~\onlinecite{Hadjighasem16_2} using a spectral clustering approach. Here, we would like to capture the surviving KAM regions as coherent structures using the level set method, as described in \Cref{alg:algorithm1}.

To identify these coherent regions, we construct the level set function $\phi$ over a uniform grid of $300\times 300$ points. The spatial domain ranges from $-2.6$ to $-0.3$ in the $x_{1}$ direction and from $-1.2$ to $1.2$ in the $x_{2}$ direction. We compute the Cauchy--Green strain tensor $C_{t0}^{t}$, with $t_{0} = 0$ and $t_{1} = 800\times 2\pi$, over the active set as the level set function evolves. Hence, the Cauchy--Green strain tensor is just computed for those grid nodes that are visited by the zero level set over its evolution.

In \cref{fig:PendulumLevelsetEvolution} (Multimedia view), we show the evolution of the zero level set toward KAM region boundaries. This example highlights how the level set method can be used for detecting multiple structures automatically.

\begin{figure}
\includegraphics[height=0.3\textwidth]{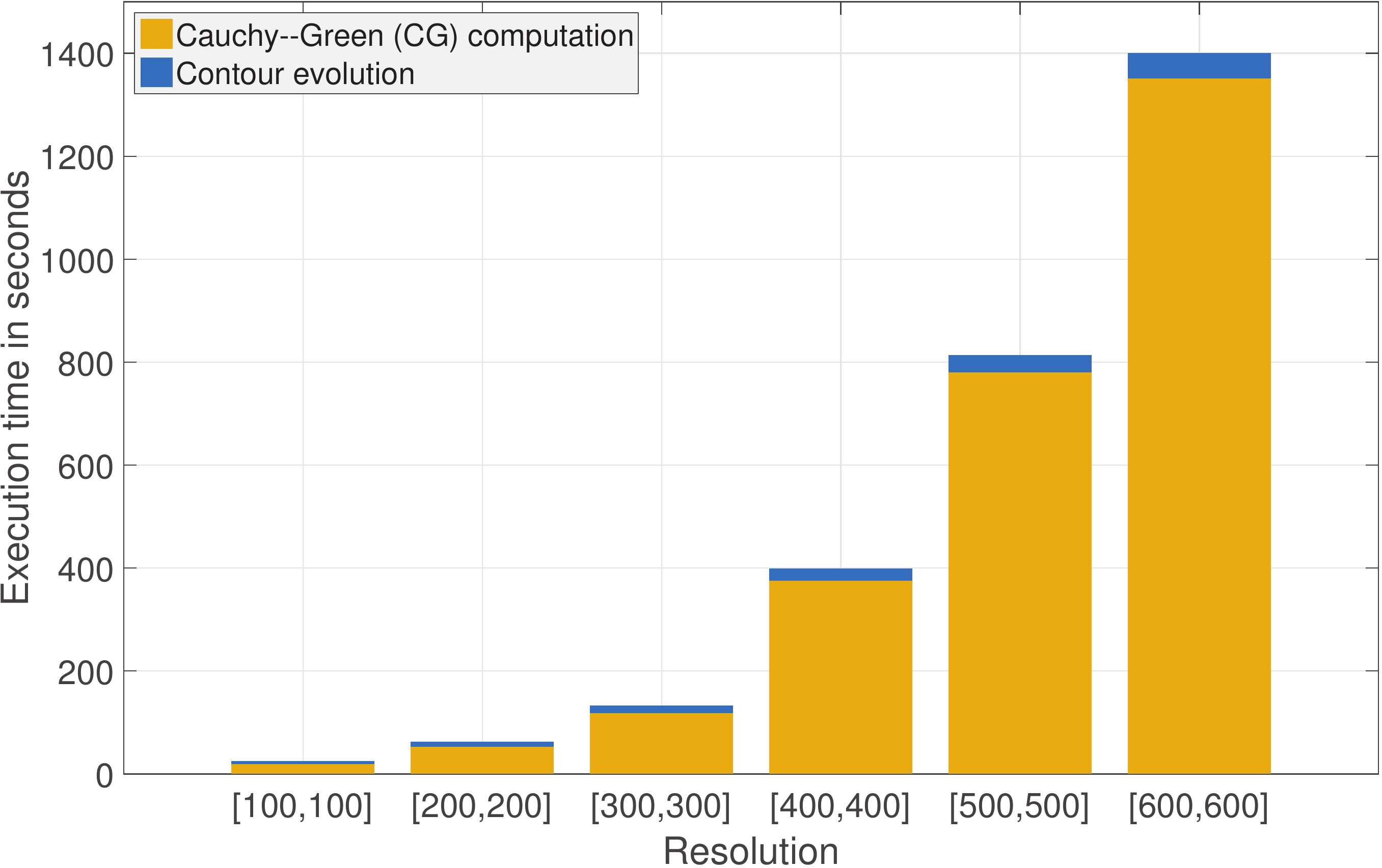}
\caption{The plot depicts the runtimes of \Cref{alg:algorithm1} for six different resolutions for the periodically forced pendulum.
The CG runtimes represent the average CPU-times for 28  MATLAB workers used in parallel in these computations. The runtimes for the contour evolution are obtained from serial computations. The computations were carried out on MATLAB R2015b installed in a computer with two 3.10 GHz Intel Xeon CPUs.}
\label{fig:PendulumRuntime}
\end{figure} 
Although \cref{fig:pendulum3} shows a good correspondence between KAM region boundaries and the zero level set, some minor discrepancies can be observed. Mainly present in the sharp corners areas, these discrepancies arise for the following reasons. First, the level set function is constructed on a uniform grid of finite resolution which can capture the sharp edges of the elliptic regions only up to a certain degree. This can be, however, enhanced using adaptive mesh generation techniques (see, e.g., Refs.~\onlinecite{Sussman99,Persson04}). Second, while the regularization term $\varepsilon\kappa$ maintains the smoothness of the interface $\Gamma$ during its evolution, it may also undesirably prevent the development of sharp corners in the evolving interface. Third, KAM tori are close to, but generally do not coincide with the infinitesimally uniformly stretching curves over a finite time interval.
\begin{figure}
\includegraphics[height=0.3\textwidth]{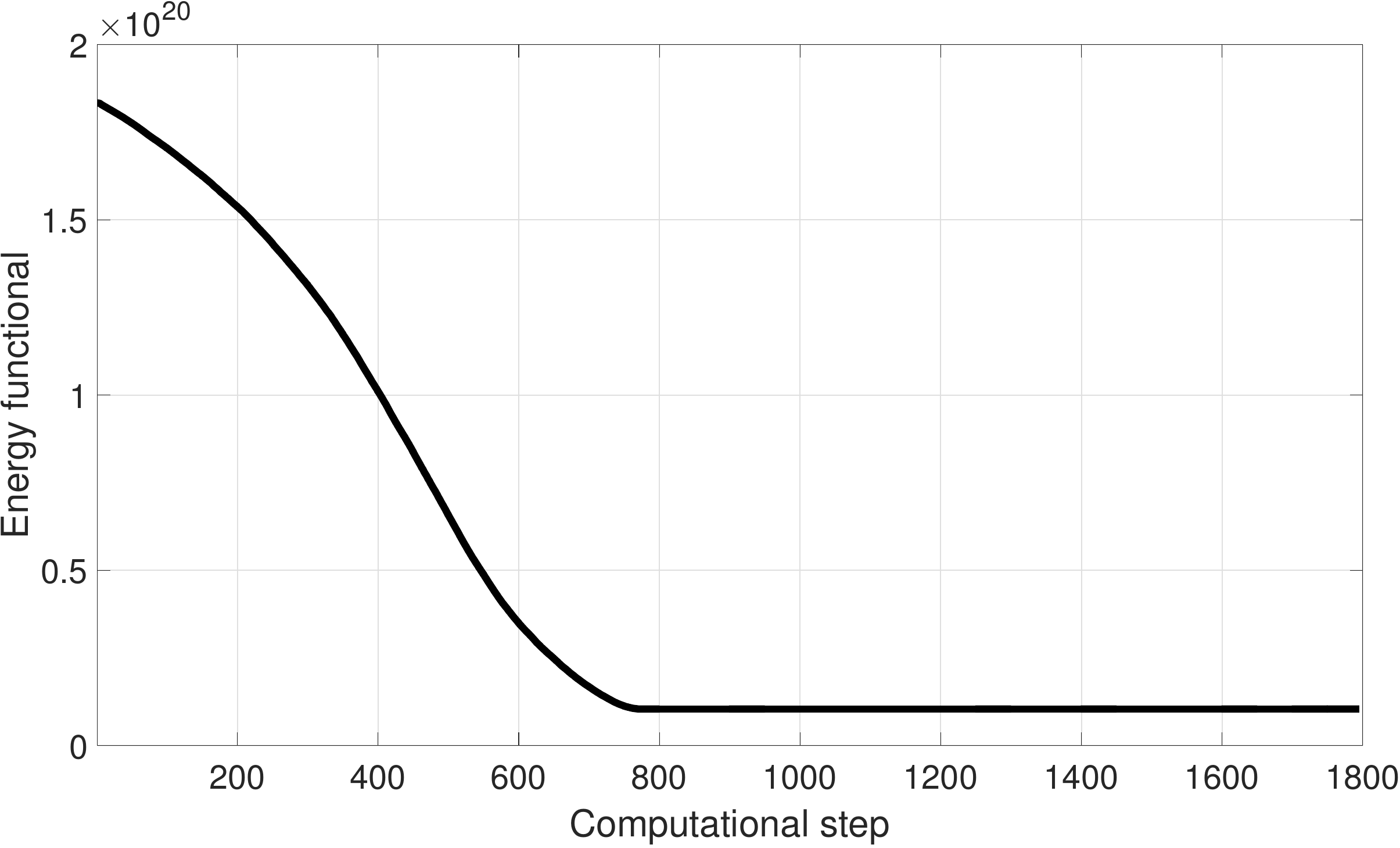}
\caption{The evolution of the energy functional vs. the computational step in a level-set optimization for the periodically forced pendulum.}
\label{fig:PendulumEnergy}
\end{figure}
\Cref{fig:PendulumRuntime} shows the execution times for two major steps of \Cref{alg:algorithm1} as a function of increasing spatial resolution of the computational domain. The main computational bottleneck, as shown in the figure, is computing the Cauchy--Green strain tensor for the active set. For this reason, we utilized parallel computing techniques with 28 MATLAB workers, with each worker just computing the CG strain tensor for a few active points. At the same time, we used simple serial computation to update the zero level set and its corresponding sparse band.

The decay of the energy functional in our numerical computation is shown \cref{fig:PendulumEnergy}. We note that the energy functional decays fast initially due to the strong non-uniform stretching present in the chaotic region.

\subsection{Jupiter's wind-velocity field}\label{section:Jupiter}
We use the level set method of \Cref{alg:algorithm1} to uncover unsteady mixing barriers in an unsteady velocity field extracted from a video footage of Jupiter's atmosphere \cite{Hadjighasem16}. The video footage is acquired over NASA's Cassini mission, covering 24 Jovian days that range from October 31 to November 9 in the year 2000. To reconstruct the velocity field, we apply the Advection Corrected Correlation Image Velocimetry (ACCIV) method \cite{Asay09} that yields a high-density, time-resolved representation of Jupiter's wind field at the cloud deck (see Ref.~\onlinecite{Hadjighasem16} for more details).

\begin{figure}
\includegraphics[height=0.26\textwidth]{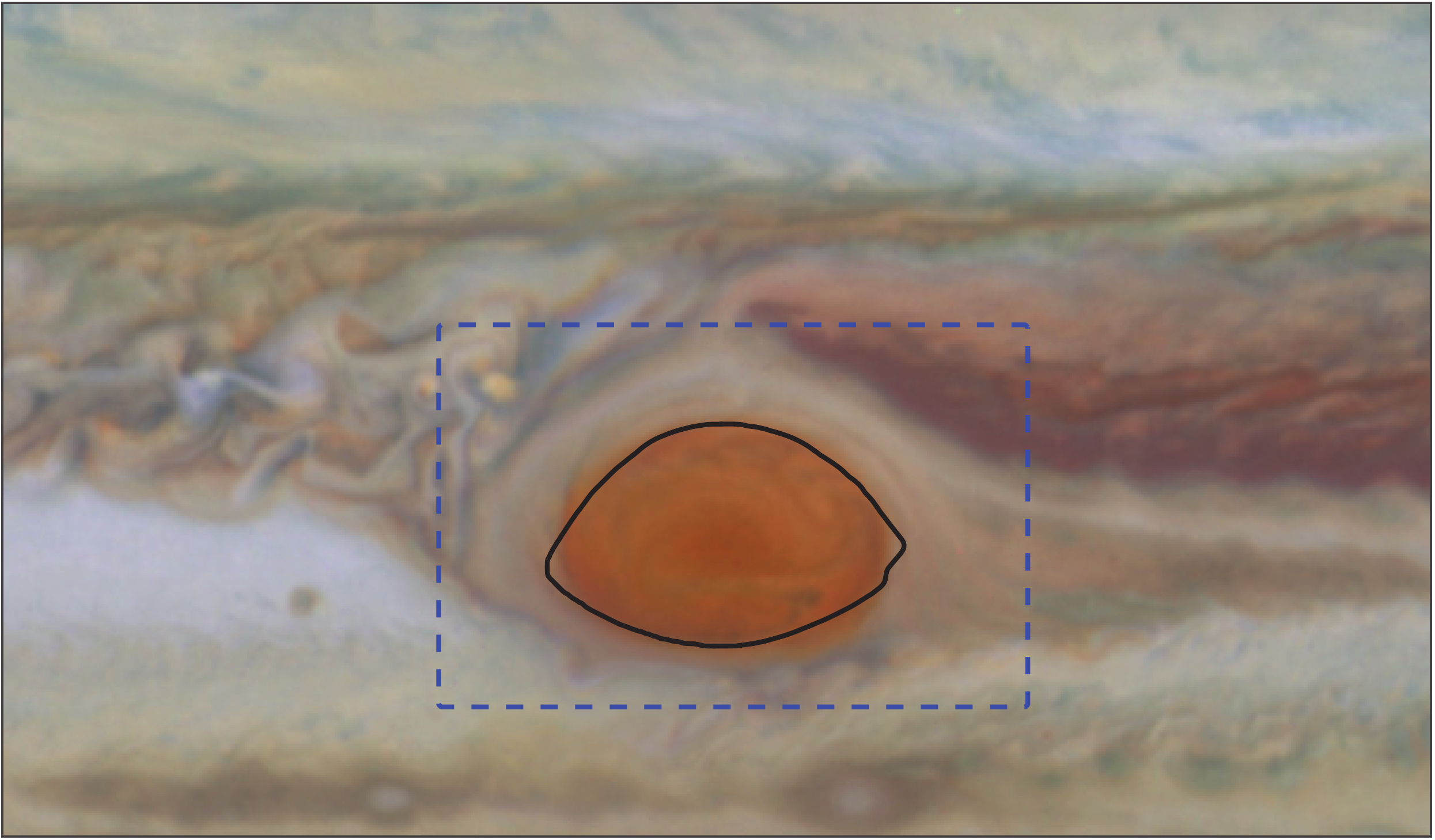}
\caption{Lagrangian vortex boundary of the GRS obtained with the level-set method shown at initial time $t_{0} = 0$. The initial zero level set is shown with blue dashed line. The new global map of Jupiter acquired by NASA's Hubble Space Telescope on January 19, 2015 is used as background. (Multimedia view)}
\label{fig:JupiterLevelsetEvolution}
\end{figure}

For the level-set computation described in \cref{section:LevelsetFormulation}, we calculate the Cauchy--Green strain tensor field $C_{t0}^{t_{1}}$, with $t_{0} = 0$ and $t_{1} = 24$ days, over a uniform grid of $300 \times 200$ points. The spatial domain $U$ ranges from $-61.6^\circ$ W to $-31.6^\circ$ W in longitude and from $-8.9^\circ$ S to $-28.9^\circ$ S in latitude. \Cref{fig:JupiterLevelsetEvolution} (Multimedia view) shows the level set-based vortex boundary for the Great Red Spot (GRS), superimposed on the cylindrical map of Jupiter acquired by  NASA's Hubble Space Telescope \cite{url_1}. 

\begin{figure}
\subfloat[\label{fig:JupiterLevelsetGeodesict0}]{\includegraphics[width=0.3\textwidth]{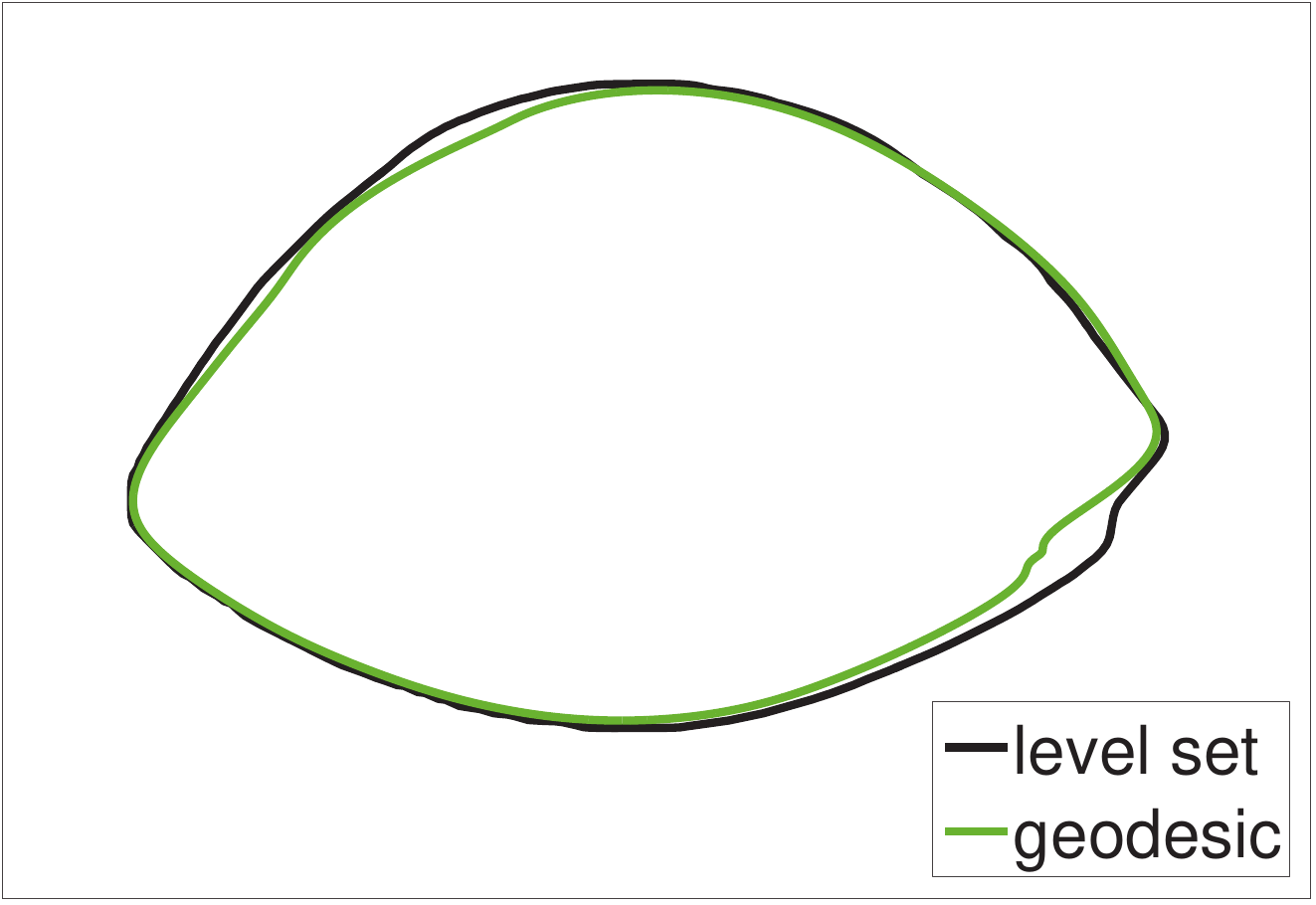}}\qquad
\subfloat[\label{fig:JupiterLevelsetGeodesictf}]{\includegraphics[width=0.3\textwidth]{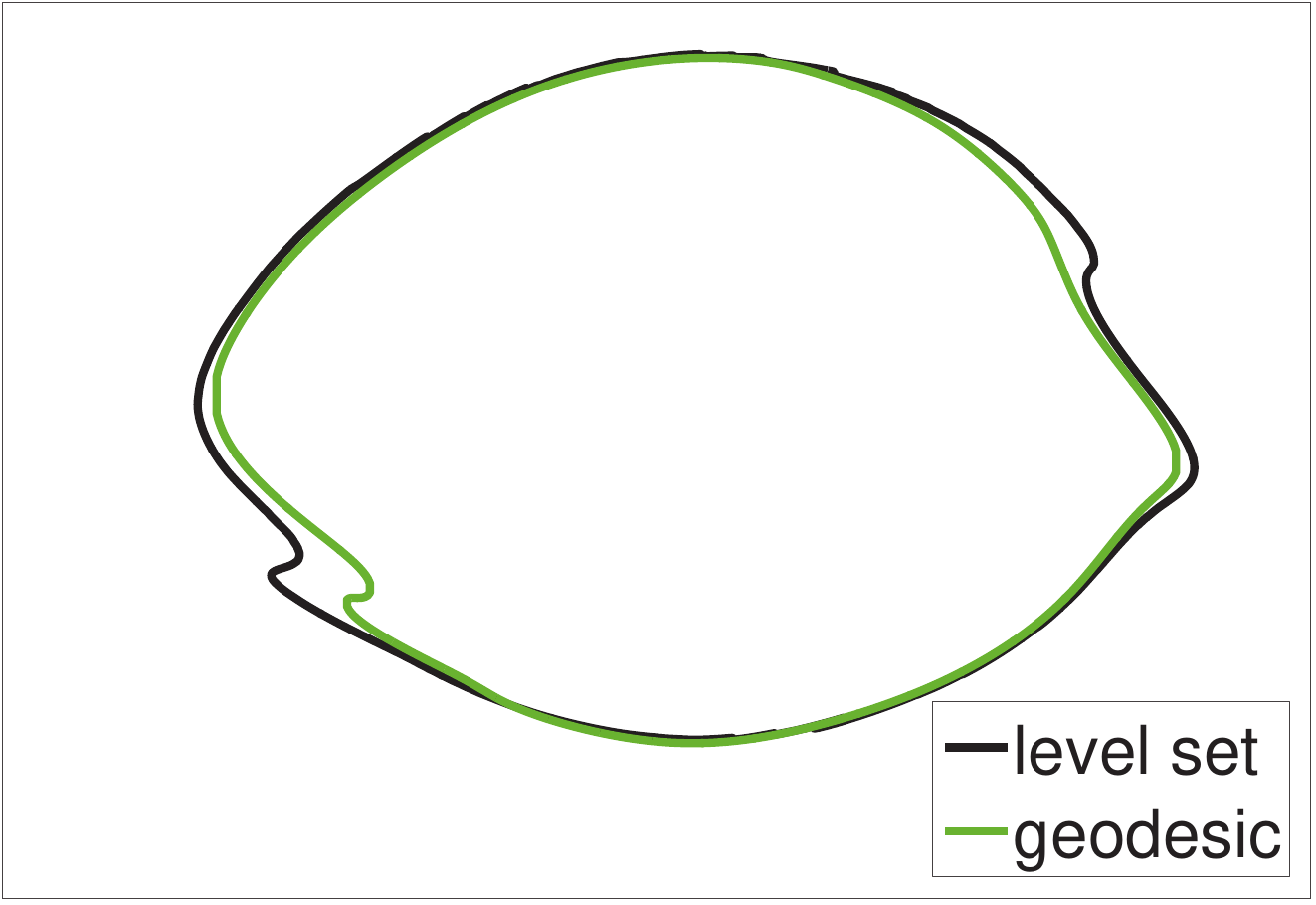}}
\caption{Geodesic vortex boundary (green) at initial time $t_{0} = 0$ for the Jupiter data set \cite{Hadjighasem16}, with the level set-based vortex boundary (black) superimposed. (b) Advected position of the Lagrangian vortex boundaries at final time $t_{1} = 24$.}
\label{fig:Jupiter_levelset_geodesic}
\end{figure}

\begin{figure}
\includegraphics[height=0.3\textwidth]{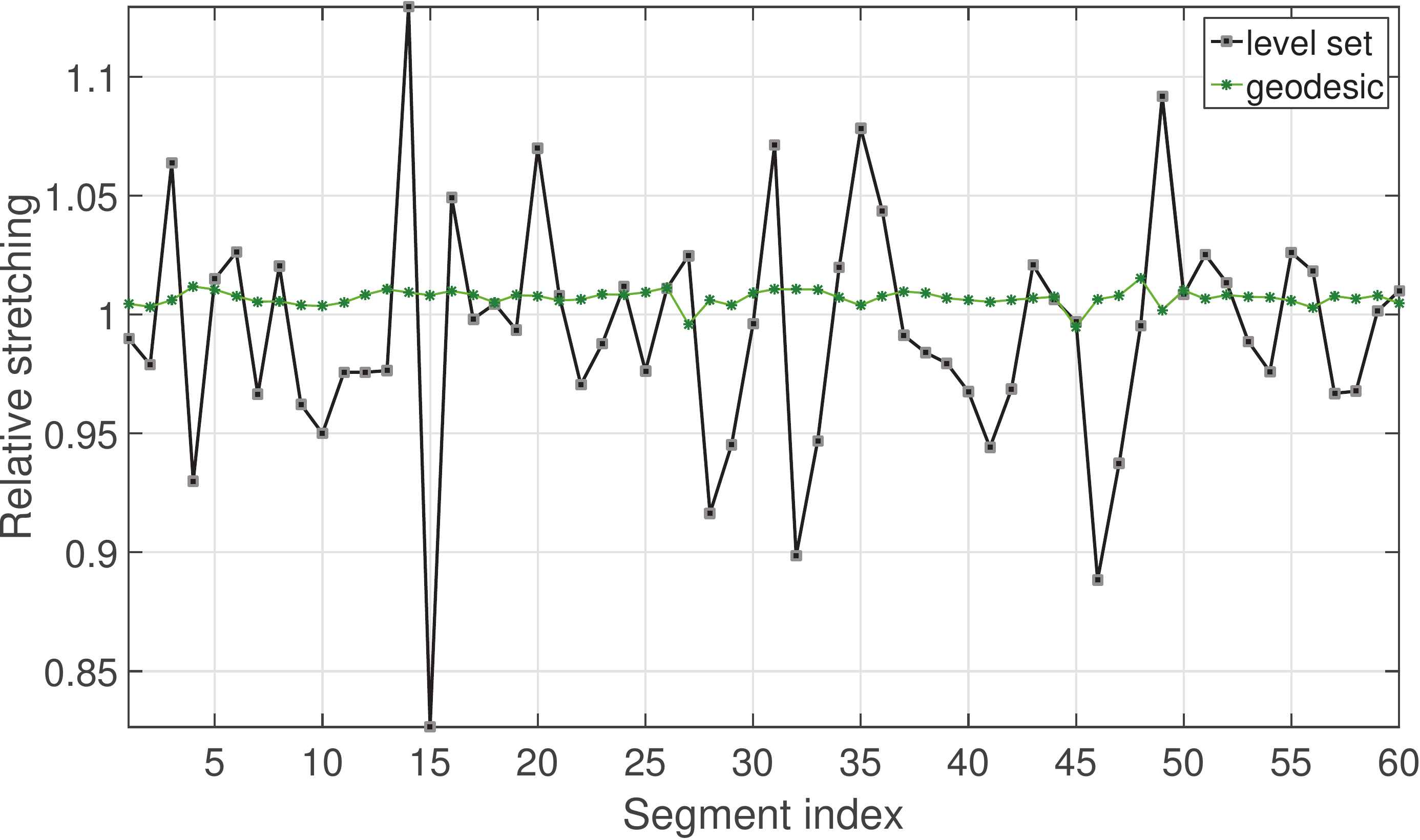}
\caption{Relative stretching of the geodesic boundary in comparison with the relative stretching of the level set-based boundary. The relative stretching of a material line segment is defined as the ratio of its length at the final time $l_{t_{1}}$ to its initial length $l_{t_{0}}$.}
\label{fig:JupiterRelativeStretching}
\end{figure}

Beyond executing \Cref{alg:algorithm1} to extract the boundary of the GRS using the level set framework, we also use this example to make a comparison with the geodesic LCS theory \cite{Haller13}. As mentioned in \cref{section:StretchingFormulation}, the latter theory seeks vortex boundaries as closed material-lines that remain perfectly non-filamenting over a finite time interval of interest. Such vortex boundaries turn out to be closed material curves in the flow that stretch uniformly by a constant factor. \Cref{fig:JupiterLevelsetGeodesict0} shows the result from the geodesic approach at the initial time \cite{Hadjighasem16}, with the level set-based vortex boundary superimposed.

\Cref{fig:JupiterLevelsetGeodesict0} shows that both methods label the GRS as a vortex, but the geodesic method yields a tighter boundary compared to the level set approach. This is because the geodesic method adopts a more stringent definition of coherence, which imposes the uniform stretching of the boundary. This observation is also consistent with the earlier comparison made between the geodesic LCS method and the more recent Lagrangian-Averaged Vorticity Deviation approach \cite{Haller16}.

In \cref{fig:JupiterLevelsetGeodesictf}, we show the advected image of the extracted vortex boundaries at the final time, confirming the sustained coherence for both boundaries over the period of 24 Jovian days. For the purposes of this comparison, we have used the numerical implementation of the geodesic eddy detection method described in Hadjighasem \& Haller \cite{Hadjighasem16}. A MATLAB implementation of this algorithm is available under \url{https://github.com/LCSETH}.

In \cref{fig:JupiterRelativeStretching}, we show a comparison of relative stretching of the geodesic vortex boundary and the level set based vortex boundary. \Cref{fig:JupiterRelativeStretching} confirms the expectation that the geodesic boundary only exhibits uniform stretching, while the level set-based boundary can exhibit larger variation in the relative stretching. The small deviation from constant stretching in the computed geodesic boundary is only due to finite sampling of the curve, as well as to the interpolation error in the computation of Cauchy--Green strain tensor field. 

While the geodesic LCS method yields a perfectly coherent boundary, the level set approach  comes with a lower computational cost for the following reasons. First, the search for a maximal limit cycle in the vector field family induced by the value of relative stretching is absent in the level set approach. Second, the evolution of the level set function is governed by a vector field which does not rely on Cauchy-Green invariants. This in turn eliminates the need for the Cauchy-Green eigendecomposition, which must be carried out with high precision close to the tensor singularities. Third, the geodesic method requires integrating a vector field for which orientational discontinuities need to be resolved locally at each integration step. Such orientational discontinuities are not present in the level set approach.

\subsection{An ocean surface data set}\label{section:Ocean}
\begin{figure*}
\subfloat[\label{fig:OceanLevelsetEvolution}]{\includegraphics[width=0.47\textwidth]{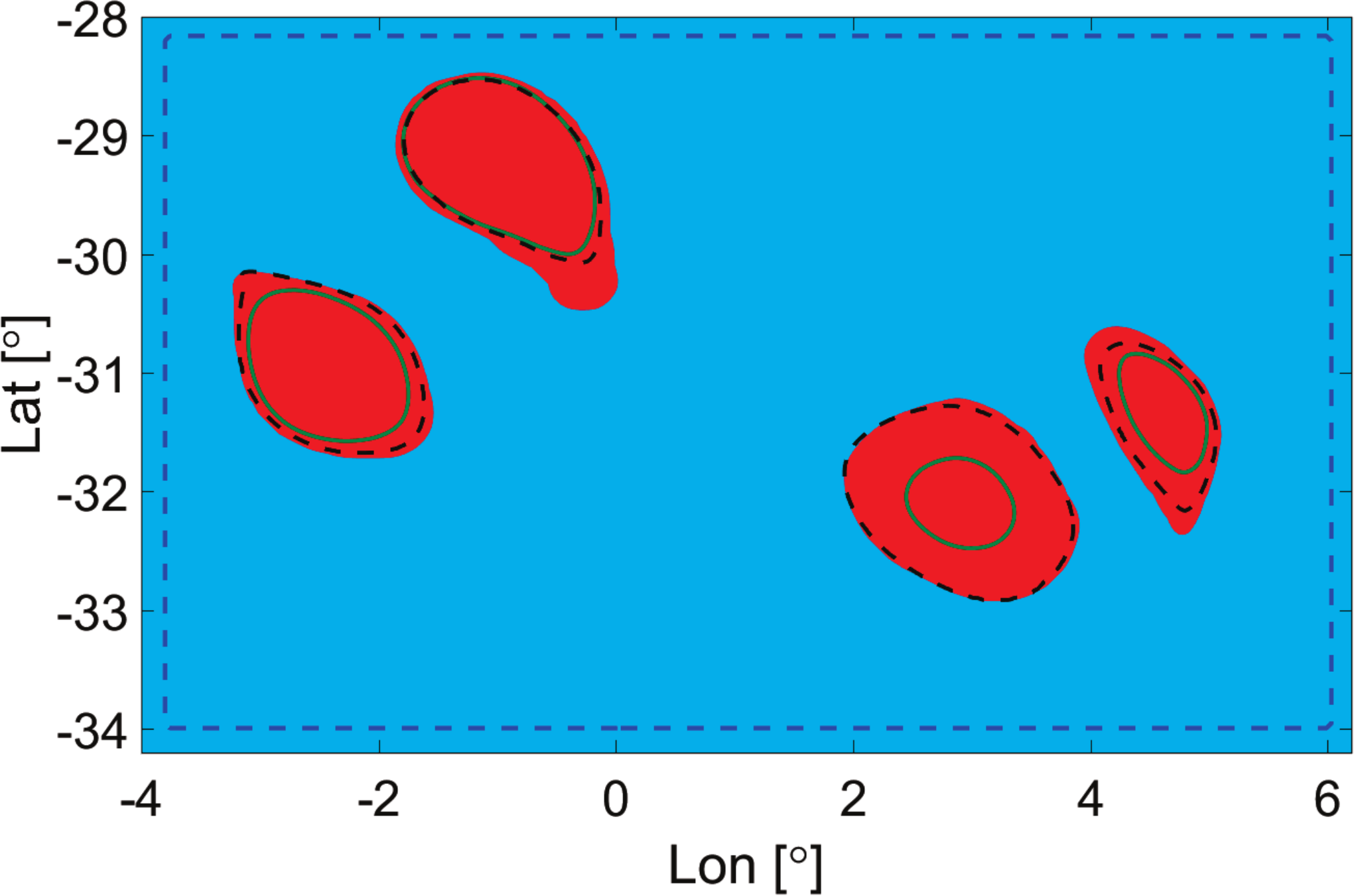}}\quad
\subfloat[\label{fig:OceanLevelsetAdvection}]{\includegraphics[width=0.47\textwidth]{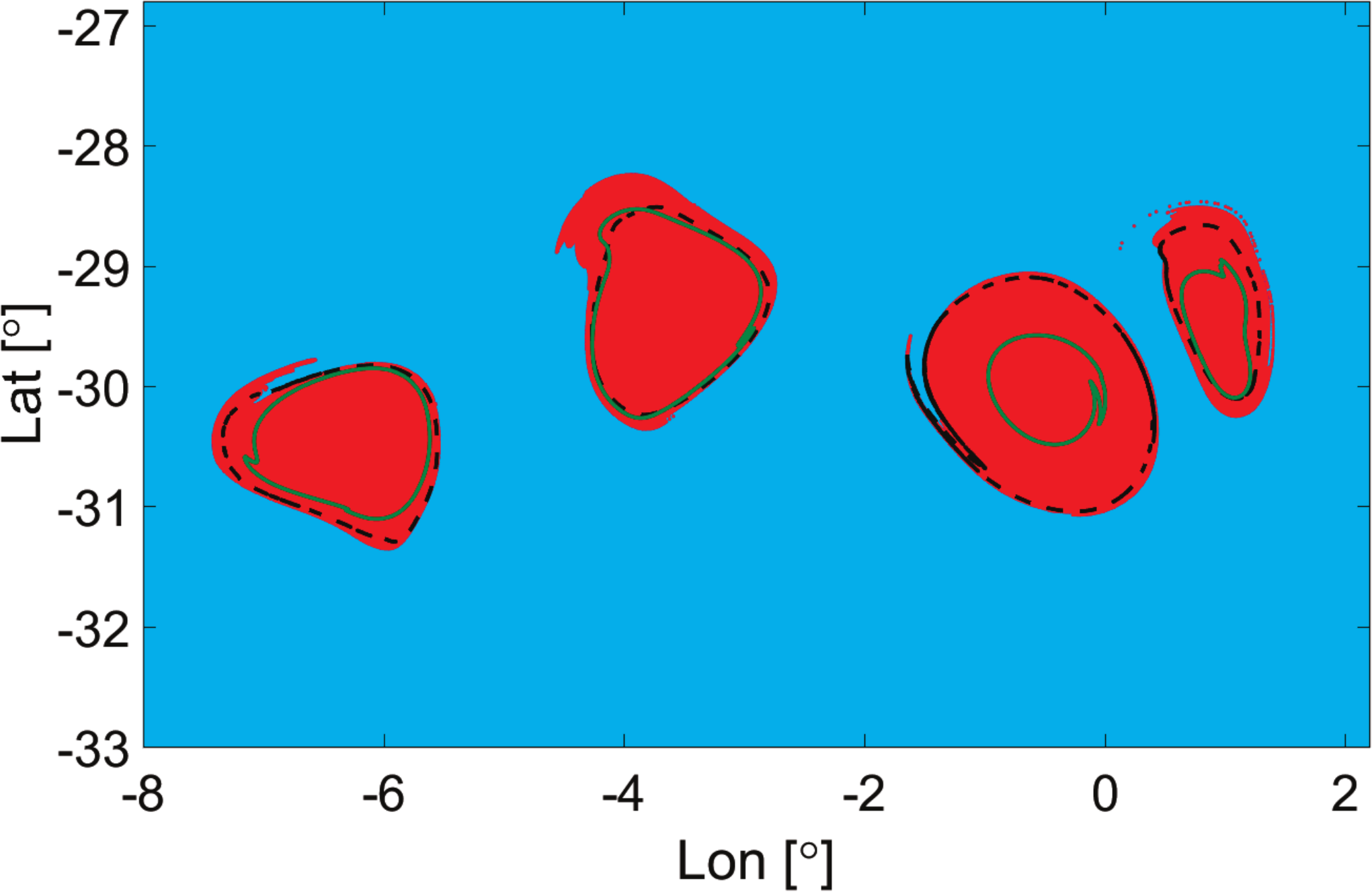}}
\caption{(a) Graph-based vortices (red) identified from \Cref{alg:algorithm2} with the stretching-based vortex boundaries (green) identified from \Cref{alg:algorithm1} at time $t_{0} =$ November 11, 2006. The initial zero level set is shown with blue dashed line. The LAVD-based vortex boundaries are shown in black for the comparison. (b) The advected positions of the vortex boundaries 90 days later at time $t_{1} =$ 9 February, 2007. (Multimedia view)}
\label{fig:Ocean}
\end{figure*}
Next, we apply \Cref{alg:algorithm2,alg:algorithm1} to a two-dimensional unsteady velocity data set derived from AVISO satellite-observed sea-surface heights (SSH) under the geostrophic approximation. In this approximation, the sea-surface height $\eta\left(\varphi,\theta,t\right)$ serves as a stream function for surface velocities in $(\varphi,\theta)$ longitude-latitude spherical coordinate system. The evolution of fluid particles is given by 
\begin{eqnarray*}
\dot{\varphi}(\varphi,\theta,t) & = & -\frac{g}{R_{E}^{2}f(\theta)\cos\theta}\ \partial_{\theta}\eta(\varphi,\theta,t),\\
\dot{\theta}(\varphi,\theta,t) & = & \frac{g}{R_{E}^{2}f(\theta)\cos\theta}\ \partial_{\varphi}\eta(\varphi,\theta,t),
\end{eqnarray*}
where $g$ is the constant of gravity, $R_E$ is the mean radius of the Earth, and $f(\theta)\equiv2\Omega_{E}\sin\theta$ is the Coriolis parameter, with $\Omega_E$ denoting the Earth's mean angular velocity.

Here, we illustate the detection of coherent Lagrangian vortices with \Cref{alg:algorithm2,alg:algorithm1} over a period of $90$ days, ranging from $t_{0} = $ November 11, 2006, to $t_{1} = $ 9 February, 2007. We select the computational domain in the longitudinal range $\left[-4^{\circ},6^{\circ}\right]$ and the latitudinal range $\left[-34^{\circ},-28^{\circ}\right]$, which falls inside the region of the Agulhas leakage in the Southern Ocean. The region in question with the same time interval is studied earlier in Ref.~\onlinecite{Haller16} using Lagrangian-Averaged Vorticity Deviation approach.

To apply \Cref{alg:algorithm2}, we select a uniform grid of $250\times 150$ points to represent the level set function $\phi$. To evolve the level set function across the active set at each iteration, we first construct a localized graph, with 64  nodes distributed uniformly in a ball of radius $R = 1/25^{\circ}$, for each active point. We then partition each localized graph into a local interior and local exterior, and find the subsequent Ncut value. The optimality of partitioning in return drives the zero level set toward the vortex boundaries (see \cref{section:GraphFormulation}). In this computation, we set the regularization term as $\varepsilon = 10^{-3}$.

\Cref{fig:OceanLevelsetEvolution} (Multimedia view) shows the time $t_{0}$ position of the vortices identified from \Cref{alg:algorithm2,alg:algorithm1}, and \cref{fig:OceanLevelsetAdvection} (Multimedia view) shows their advected positions at time $t_{1}$, confirming the coherence of the extracted vortices over the 90-day period. The LAVD-based vortex boundaries are also shown in black for the comparison.

As shown in \cref{fig:Ocean}, the results of \Cref{alg:algorithm1} and \Cref{alg:algorithm2} can differ from each other as they are based on different coherence principles. In fact, each of \Cref{alg:algorithm1,alg:algorithm2} has its own advantages and disadvantages. For instance, \Cref{alg:algorithm1} compared to \Cref{alg:algorithm2} use a more stringent notion of coherence which usually results in smaller vortex boundaries (see \cref{fig:Ocean}). However, \Cref{alg:algorithm2} is computationally more expensive than \Cref{alg:algorithm1}. The main reason for presenting both algorithms is to emphasize that the level set methodology can be used for reformulating different approaches developed for detecting coherent structures in the fluid flows.

\section{Conclusion}
We have demonstrated the application of the variational level set methodology to coherent material vortex detection in fluid flows. To identify coherent structures, we minimize appropriate energy functionals defining the boundaries of coherent vortices. We carry out the minimization via a gradient-descent method, that drives the zero level set towards the desired boundaries.

We have illustrated the performance of the proposed technique on two different energy functionals, each using a different Lagrangian notion of coherence. Our first variational formulation seeks coherent vortices as closed material lines that are close to uniformly stretching. This notion of coherence derives from earlier work of Haller \& Beron-Vera \cite{Haller13}. We show the effectiveness of the corresponding approach by detecting Lagrangian coherent vortices in periodic and unsteady two-dimensional flows.

In the second approach, we adopt the idea of normalized graph cut \cite{Shi00} to identify coherent structures based on the proximity of particles in the spatio-temporal domain. Here, we conceive coherent structures, in a fashion similar to Refs.~\onlinecite{Hadjighasem16_2,Froyland15}, as a set of Lagrangian particles that remain tightly grouped. We apply this second approach in our last example, the ocean surface data set, to identify Agulhas eddies in the Southern Ocean.

A drawback of the level set technique is the effort required for the construction of energy  functionals whose local minima mark the vortex boundaries. A reward for this effort is a versatile numerical platform that can capture vortices in an automated fashion.

Future challenges include extending the current level set approach to three-dimensional problems and using parallel implementation for speeding up the related calculations.

\section*{Acknowledgements}
The authors would like to thank Mattia Serra and David Oettinger for helpful discussions on the subject of this paper. The altimeter products used in this work are produced by SSALTO/DUACS and distributed by AVISO, with support from CNES.

\appendix

\section{Numerical aspects}\label[appendix]{appendix:Implementation}
The numerical implementation of Equations \eqref{eq:functional_derivative} and \eqref{eq:graph_fun_der} is simple, but requires some care to ensure sufficient accuracy and efficiency. In this section, we address these implementation aspects.

\begin{description}
\item[Stability and CFL condition] To keep numerical stability and obtain accurate approximation results, the time step for solving \eqref{eq:LevelSetEvolution} with explicit time-marching scheme must satisfy the \emph{Courant-Friedrichs-Lewy} (CFL) condition \cite{Osher06}, which states the front should not cross more than one grid cell at each time step:
\begin{equation*}
\Delta \tau\left( \frac{\max\left| u \right|}{\Delta x} \right) = \mu, \quad 0<\mu<1.
\end{equation*}
Here, $u$ refers to the speed with which the zero level set propagates. A common near-optimal choice for the \emph{CFL number} is $\mu = 0.9$, and a common
conservative choice is $\mu = 0.5$ (cf. Ref.~\onlinecite{Osher06}). 

For stability concerns, \emph{implicit} or \emph{semi-implicit} methods may also improve the efficiency of level set methods. Compared to the time steps of explicit schemes limited by a CFL condition, the implicit or semi-implicit level set methods allow for larger time steps (see, for example, Ref.~\onlinecite{You15}). Consequently, the convergence of implicit or semi-implicit schemes is usually faster compared to the explicit methods.

\item[Reinitialization] In general, even if we initialize the level set function $\phi$ as a signed distance function, it is not guaranteed to remain a distance function at later times. As a consequence, the level set function $\phi$ develops steep or flat shapes during the evolution, making the results inaccurate. Classic level set methods often use the re-initialization remedy to avoid this problem, that is, periodically initialize the level set function as a signed distance function using either the fast marching method \cite{Sethian96} or PDE-based approaches \cite{Sussman94}. The re-initialization process, however, is complicated, expensive and has an unwanted side effect of shifting the zero level set away from its original location \cite{Li05}. Moreover, this process is conducted in an ad-hoc manner because there is no rule as to when and how to reinitialize the level set function to a signed distance function. A better approach is to limit re-initialization \cite{Sethian01} or use methods that do not require re-initialization at all (see Refs.~\onlinecite{Li05,Zhang13} for examples).

\item[Finite Difference Scheme]
Equation \eqref{eq:functional_derivative} is a nonlinear Hamilton-Jacobi equation composed of  both parabolic and hyperbolic terms. When implementing Eq. \eqref{eq:functional_derivative}, one must give special attention to how parabolic terms, such as $|\nabla\phi|$, are calculated, as standard finite difference methods fail for non-linear hyperbolic PDEs. Thus, one needs the special machinery of \emph{upwind finite differencing} or \emph{upwinding}, where spatial derivatives are computed using one-sided differencing based on the direction of propagation. We make use of the state-of-the-art high order ENO \cite{Osher88,Osher91} and WENO \cite{Jiang00} schemes in our implementations, whenever it is appropriate to do so.

\item[Level Set Regularization] In \cref{section:background}, we assumed that the interface $\Gamma$ stays smooth over its evolution, but in applications, smoothness is often lost. A well-known example is the cosine curve evolving with unit speed $u(x,\tau) = 1$, where the propagating curve develops a sharp corner in finite
time \cite{Sethian01}.  Once the corner develops, the normal direction becomes undefined and the differentiability of the interface is lost. Thus, it is important to ensure that the interface $\Gamma$ stays smooth and non-intersecting all along its evolution. This is commonly achieved by adding a regularization term $\varepsilon\kappa$ to the evolution equation (see Refs.~\onlinecite{Osher88,Sethian01,You15} for examples). The curvature term $\varepsilon\kappa$ regularizes the interface by accelerating the movement of those segments of the interface that remain behind the average speed of the interface and slowing down the segments marching faster than the average speed. The parameter $\varepsilon$ determines the strength of regularization. If $\varepsilon$ is large, the regularization term will smooth out interface irregularities such that the interface ultimately will become convex. If $\varepsilon$ is small, the front will maintain sharp curvatures and may have a concave geometry at the end of the evolution.

\item[Narrow Band] The classic level set approach evolves the level set function $\phi$ by solving an initial value problem for a partial differential equation in the entire computational domain. This is superfluous if only information near the zero level set is of interest. Instead, an efficient modification is to perform the computation in a neighborhood or \emph{narrow band} of the zero level set, as introduced by Adalsteinsson and Sethian \cite{Adalsteinsson95}.  The idea of the narrow band approach was later extended to the \emph{Sparse Field Method} (SFM), in whih the narrow band is only one pixel wide and the level set function is re-initialized with a distance transform in each iteration \cite{Whitaker98}. We will discuss the Sparse Field Method further in \cref{appendix:SFM}.
\end{description}

More details concerning the numerical schemes for level set methods can be found in Ref.~\onlinecite{Osher06}.

\section{Sparse Field Method}
\begin{figure}
\subfloat[\label{fig:SFM}]{\includegraphics[height=0.3\textwidth]{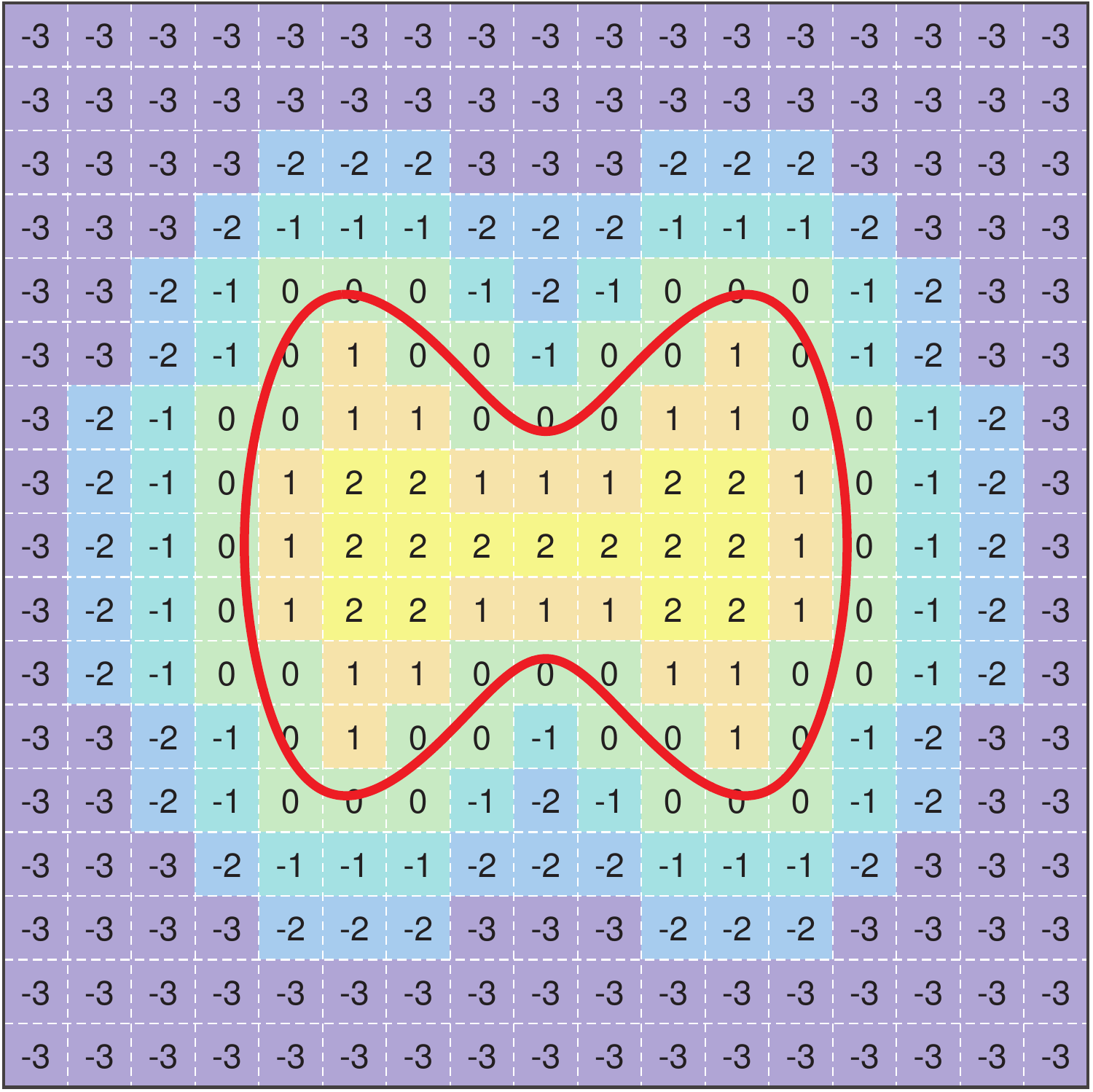}}\quad
\subfloat[\label{fig:SFMLists}]{\includegraphics[height=0.3\textwidth]{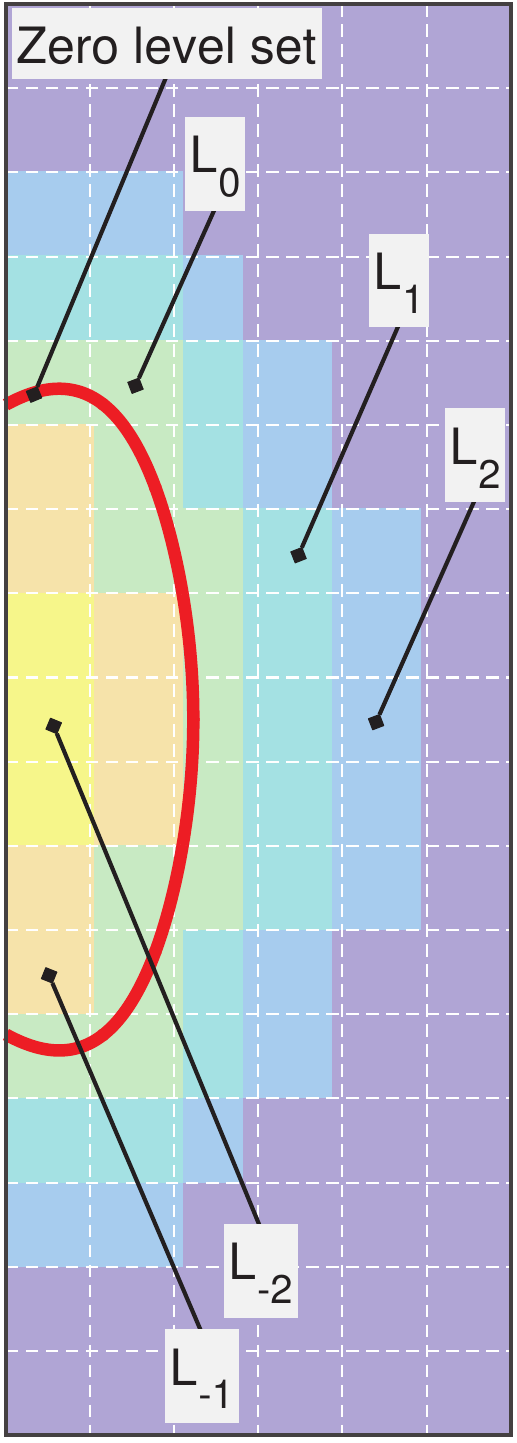}}
\caption{(a) One example of the initialization in SFM. A zero level curve of a 2D scalar field passes through a finite set of cells. Only those cells nearest to the level curve are relevant to the evolution of that curve. (b) Visual representation of the neighborhood layers.}
\label{fig:SparseFieldMethod}
\end{figure}

\label[appendix]{appendix:SFM}
In classical level set methods, the value of the level set function $\phi$ is updated in the full computational domain, which is computationally costly.  Narrow band methods \cite{Adalsteinsson95,Tsai03} address this problem by only updating pixels near the evolving curve. To optimize and simplify the implementation of the narrow-band scheme, Whitaker \cite{Whitaker98} proposed the Sparse Field Method which takes the narrow-band strategy to the extreme. The basic idea of the SFM is to use lists of points that represent the zero level set as well as points adjacent to the zero level set (see \cref{fig:SparseFieldMethod}). By using these lists and carefully adding and removing points from the appropriate list, the level set function $\phi$ can efficiently be maintained. The fact that the SFM uses lists to keep track of the points near the zero level set means that the computational speed at each iteration depends only on the length of the curve $|\Gamma|$, and not on the size of the domain. 

We call the minimal connected set of grid points that are closest to the level set as the \emph{active set}, denoting it by $L_{0}$, and the individual elements in this set are the \emph{active points}.  We then define its neighborhood layers by $L_{\pm i}$ for $i=\pm 1,\ldots,\pm N$, where $i$ indicates the city block distance of a neighborhood point from the nearest active point (see \cref{fig:SFMLists}). In this paper, we use up to the second-order derivatives of $\phi$, so we need only five layers: $L_{2}$, $L_{1}$, $L_{0}$, $L_{-1}$, and $L_{-2}$. In addition to the lists, two arrays are used to save the information of the above lists. The first is the $\phi$ array which has the same dimensions as the computational domain and should be stored at full floating point precision. The second array is a label map which is used to record the status of each point and takes integer values $\lbrace-3,-2,-1,0,1,2,3\rbrace$, as shown in \cref{fig:SFM}.

The procedure of SFM can be divided into three main steps: initialization, curve evolution and updating the lists. The initialization process of the interface is fairly simple and starts by defining a level set function whose zero level set is explicitly stored at various grid points. This is done by assigning the corresponding points in  $\phi$ to $0$, and by adding them to the $L_{0}$ list. The other lists are then filled with points according to their distance from the nearest active point, and are updated accordingly. Next, points in $\phi$ that are members of the active set $L_{0}$ are updated by the level set evolution equation. These changes are then reflected in the neighboring layers with the simple numerical procedure specified in Ref.~\onlinecite{Whitaker98}, and the lists are updated accordingly. How these steps are executed is described in details in Refs.~\onlinecite{Whitaker98,Lankton09}.

\bibliography{Level_Set_lib}

\begin{thebibliography}{47}%
\makeatletter
\providecommand \@ifxundefined [1]{%
 \@ifx{#1\undefined}
}%
\providecommand \@ifnum [1]{%
 \ifnum #1\expandafter \@firstoftwo
 \else \expandafter \@secondoftwo
 \fi
}%
\providecommand \@ifx [1]{%
 \ifx #1\expandafter \@firstoftwo
 \else \expandafter \@secondoftwo
 \fi
}%
\providecommand \natexlab [1]{#1}%
\providecommand \enquote  [1]{``#1''}%
\providecommand \bibnamefont  [1]{#1}%
\providecommand \bibfnamefont [1]{#1}%
\providecommand \citenamefont [1]{#1}%
\providecommand \href@noop [0]{\@secondoftwo}%
\providecommand \href [0]{\begingroup \@sanitize@url \@href}%
\providecommand \@href[1]{\@@startlink{#1}\@@href}%
\providecommand \@@href[1]{\endgroup#1\@@endlink}%
\providecommand \@sanitize@url [0]{\catcode `\\12\catcode `\$12\catcode
  `\&12\catcode `\#12\catcode `\^12\catcode `\_12\catcode `\%12\relax}%
\providecommand \@@startlink[1]{}%
\providecommand \@@endlink[0]{}%
\providecommand \url  [0]{\begingroup\@sanitize@url \@url }%
\providecommand \@url [1]{\endgroup\@href {#1}{\urlprefix }}%
\providecommand \urlprefix  [0]{URL }%
\providecommand \Eprint [0]{\href }%
\providecommand \doibase [0]{http://dx.doi.org/}%
\providecommand \selectlanguage [0]{\@gobble}%
\providecommand \bibinfo  [0]{\@secondoftwo}%
\providecommand \bibfield  [0]{\@secondoftwo}%
\providecommand \translation [1]{[#1]}%
\providecommand \BibitemOpen [0]{}%
\providecommand \bibitemStop [0]{}%
\providecommand \bibitemNoStop [0]{.\EOS\space}%
\providecommand \EOS [0]{\spacefactor3000\relax}%
\providecommand \BibitemShut  [1]{\csname bibitem#1\endcsname}%
\let\auto@bib@innerbib\@empty
\bibitem [{\citenamefont {Haller}\ and\ \citenamefont
  {Beron-Vera}(2013)}]{Haller13}%
  \BibitemOpen
  \bibfield  {author} {\bibinfo {author} {\bibfnamefont {G.}~\bibnamefont
  {Haller}}\ and\ \bibinfo {author} {\bibfnamefont {F.~J.}\ \bibnamefont
  {Beron-Vera}},\ }\bibfield  {title} {\enquote {\bibinfo {title} {{Coherent
  Lagrangian vortices: the black holes of turbulence}},}\ }\href {\doibase
  10.1017/jfm.2013.391} {\bibfield  {journal} {\bibinfo  {journal} {Journal of
  Fluid Mechanics}\ }\textbf {\bibinfo {volume} {731}},\ \bibinfo {pages} {R4
  (10 pages)} (\bibinfo {year} {2013})}\BibitemShut {NoStop}%
\bibitem [{\citenamefont {Hadjighasem}\ \emph {et~al.}(2016)\citenamefont
  {Hadjighasem}, \citenamefont {Karrasch}, \citenamefont {Teramoto},\ and\
  \citenamefont {Haller}}]{Hadjighasem16_2}%
  \BibitemOpen
  \bibfield  {author} {\bibinfo {author} {\bibfnamefont {A.}~\bibnamefont
  {Hadjighasem}}, \bibinfo {author} {\bibfnamefont {D.}~\bibnamefont
  {Karrasch}}, \bibinfo {author} {\bibfnamefont {H.}~\bibnamefont {Teramoto}},
  \ and\ \bibinfo {author} {\bibfnamefont {G.}~\bibnamefont {Haller}},\
  }\bibfield  {title} {\enquote {\bibinfo {title} {{Spectral-clustering
  approach to Lagrangian vortex detection}},}\ }\href {\doibase
  10.1103/PhysRevE.93.063107} {\bibfield  {journal} {\bibinfo  {journal} {Phys.
  Rev. E}\ }\textbf {\bibinfo {volume} {93}},\ \bibinfo {pages} {063107}
  (\bibinfo {year} {2016})}\BibitemShut {NoStop}%
\bibitem [{\citenamefont {Haller}(2015)}]{Haller15}%
  \BibitemOpen
  \bibfield  {author} {\bibinfo {author} {\bibfnamefont {G.}~\bibnamefont
  {Haller}},\ }\bibfield  {title} {\enquote {\bibinfo {title} {{Lagrangian
  Coherent Structures}},}\ }\href {\doibase
  10.1146/annurev-fluid-010313-141322} {\bibfield  {journal} {\bibinfo
  {journal} {Annual Review of Fluid Mechanics}\ }\textbf {\bibinfo {volume}
  {47}},\ \bibinfo {pages} {137--162} (\bibinfo {year} {2015})}\BibitemShut
  {NoStop}%
\bibitem [{\citenamefont {Hadjighasem}\ and\ \citenamefont
  {Haller}(2016)}]{Hadjighasem16}%
  \BibitemOpen
  \bibfield  {author} {\bibinfo {author} {\bibfnamefont {A.}~\bibnamefont
  {Hadjighasem}}\ and\ \bibinfo {author} {\bibfnamefont {G.}~\bibnamefont
  {Haller}},\ }\bibfield  {title} {\enquote {\bibinfo {title} {{Geodesic
  Transport Barriers in Jupiter's Atmosphere: A Video-Based Analysis}},}\
  }\href {\doibase 10.1137/140983665} {\bibfield  {journal} {\bibinfo
  {journal} {SIAM Review}\ }\textbf {\bibinfo {volume} {58}},\ \bibinfo {pages}
  {69--89} (\bibinfo {year} {2016})}\BibitemShut {NoStop}%
\bibitem [{\citenamefont {Haller}\ \emph {et~al.}(2016)\citenamefont {Haller},
  \citenamefont {Hadjighasem}, \citenamefont {Farazmand},\ and\ \citenamefont
  {Huhn}}]{Haller16}%
  \BibitemOpen
  \bibfield  {author} {\bibinfo {author} {\bibfnamefont {G.}~\bibnamefont
  {Haller}}, \bibinfo {author} {\bibfnamefont {A.}~\bibnamefont {Hadjighasem}},
  \bibinfo {author} {\bibfnamefont {M.}~\bibnamefont {Farazmand}}, \ and\
  \bibinfo {author} {\bibfnamefont {F.}~\bibnamefont {Huhn}},\ }\bibfield
  {title} {\enquote {\bibinfo {title} {{Defining coherent vortices objectively
  from the vorticity}},}\ }\href {\doibase 10.1017/jfm.2016.151} {\bibfield
  {journal} {\bibinfo  {journal} {Journal of Fluid Mechanics}\ }\textbf
  {\bibinfo {volume} {795}},\ \bibinfo {pages} {136--173} (\bibinfo {year}
  {2016})}\BibitemShut {NoStop}%
\bibitem [{\citenamefont {Farazmand}\ and\ \citenamefont
  {Haller}(2016)}]{Farazmand15}%
  \BibitemOpen
  \bibfield  {author} {\bibinfo {author} {\bibfnamefont {M.}~\bibnamefont
  {Farazmand}}\ and\ \bibinfo {author} {\bibfnamefont {G.}~\bibnamefont
  {Haller}},\ }\bibfield  {title} {\enquote {\bibinfo {title} {{Polar rotation
  angle identifies elliptic islands in unsteady dynamical systems}},}\ }\href
  {\doibase http://dx.doi.org/10.1016/j.physd.2015.09.007} {\bibfield
  {journal} {\bibinfo  {journal} {Physica D: Nonlinear Phenomena}\ }\textbf
  {\bibinfo {volume} {315}},\ \bibinfo {pages} {1 -- 12} (\bibinfo {year}
  {2016})}\BibitemShut {NoStop}%
\bibitem [{\citenamefont {Froyland}\ and\ \citenamefont
  {Padberg-Gehle}(2014)}]{Froyland14}%
  \BibitemOpen
  \bibfield  {author} {\bibinfo {author} {\bibfnamefont {G.}~\bibnamefont
  {Froyland}}\ and\ \bibinfo {author} {\bibfnamefont {K.}~\bibnamefont
  {Padberg-Gehle}},\ }\bibfield  {title} {\enquote {\bibinfo {title} {{Chapter
  9 Almost-Invariant and Finite-Time Coherent Sets: Directionality, Duration,
  and Diffusion}},}\ }\href {\doibase 10.1007/978-1-4939-0419-8__9} {\bibfield
  {journal} {\bibinfo  {journal} {Ergodic Theory, Open Dynamics, and Coherent
  Structures}\ }\textbf {\bibinfo {volume} {70}},\ \bibinfo {pages} {171--216}
  (\bibinfo {year} {2014})}\BibitemShut {NoStop}%
\bibitem [{\citenamefont {Budi\v{s}i\'{c}}\ and\ \citenamefont
  {Mezi\'{c}}(2012)}]{Budisic12_1}%
  \BibitemOpen
  \bibfield  {author} {\bibinfo {author} {\bibfnamefont {M.}~\bibnamefont
  {Budi\v{s}i\'{c}}}\ and\ \bibinfo {author} {\bibfnamefont {I.}~\bibnamefont
  {Mezi\'{c}}},\ }\bibfield  {title} {\enquote {\bibinfo {title} {{Geometry of
  the ergodic quotient reveals coherent structures in flows}},}\ }\href
  {\doibase 10.1016/j.physd.2012.04.006} {\bibfield  {journal} {\bibinfo
  {journal} {Physica D-Nonlinear Phenomena}\ }\textbf {\bibinfo {volume}
  {241}},\ \bibinfo {pages} {1255--1269} (\bibinfo {year} {2012})}\BibitemShut
  {NoStop}%
\bibitem [{\citenamefont {Allshouse}\ and\ \citenamefont
  {Thiffeault}(2012)}]{Allshouse12}%
  \BibitemOpen
  \bibfield  {author} {\bibinfo {author} {\bibfnamefont {M.~R.}\ \bibnamefont
  {Allshouse}}\ and\ \bibinfo {author} {\bibfnamefont {J.-L.}\ \bibnamefont
  {Thiffeault}},\ }\bibfield  {title} {\enquote {\bibinfo {title} {Detecting
  coherent structures using braids},}\ }\href {\doibase
  http://dx.doi.org/10.1016/j.physd.2011.10.002} {\bibfield  {journal}
  {\bibinfo  {journal} {Physica D: Nonlinear Phenomena}\ }\textbf {\bibinfo
  {volume} {241}},\ \bibinfo {pages} {95 -- 105} (\bibinfo {year}
  {2012})}\BibitemShut {NoStop}%
\bibitem [{\citenamefont {Froyland}\ and\ \citenamefont
  {Padberg-Gehle}(2015)}]{Froyland15}%
  \BibitemOpen
  \bibfield  {author} {\bibinfo {author} {\bibfnamefont {G.}~\bibnamefont
  {Froyland}}\ and\ \bibinfo {author} {\bibfnamefont {K.}~\bibnamefont
  {Padberg-Gehle}},\ }\bibfield  {title} {\enquote {\bibinfo {title} {{A
  rough-and-ready cluster-based approach for extracting finite-time coherent
  sets from sparse and incomplete trajectory data}},}\ }\href {\doibase
  10.1063/1.4926372} {\bibfield  {journal} {\bibinfo  {journal} {Chaos}\
  }\textbf {\bibinfo {volume} {25}},\ \bibinfo {pages} {087406} (\bibinfo
  {year} {2015})}\BibitemShut {NoStop}%
\bibitem [{\citenamefont {Rypina}\ \emph {et~al.}(2011)\citenamefont {Rypina},
  \citenamefont {Scott}, \citenamefont {Pratt},\ and\ \citenamefont
  {Brown}}]{Rypina11}%
  \BibitemOpen
  \bibfield  {author} {\bibinfo {author} {\bibfnamefont {I.~I.}\ \bibnamefont
  {Rypina}}, \bibinfo {author} {\bibfnamefont {S.~E.}\ \bibnamefont {Scott}},
  \bibinfo {author} {\bibfnamefont {L.~J.}\ \bibnamefont {Pratt}}, \ and\
  \bibinfo {author} {\bibfnamefont {M.~G.}\ \bibnamefont {Brown}},\ }\bibfield
  {title} {\enquote {\bibinfo {title} {{Investigating the connection between
  complexity of isolated trajectories and Lagrangian coherent structures}},}\
  }\href {\doibase 10.5194/npg-18-977-2011} {\bibfield  {journal} {\bibinfo
  {journal} {Nonlinear Processes in Geophysics}\ }\textbf {\bibinfo {volume}
  {18}},\ \bibinfo {pages} {977--987} (\bibinfo {year} {2011})}\BibitemShut
  {NoStop}%
\bibitem [{\citenamefont {Mezi{\'c}}\ \emph {et~al.}(2010)\citenamefont
  {Mezi{\'c}}, \citenamefont {Loire}, \citenamefont {Fonoberov},\ and\
  \citenamefont {Hogan}}]{Mezic10}%
  \BibitemOpen
  \bibfield  {author} {\bibinfo {author} {\bibfnamefont {I.}~\bibnamefont
  {Mezi{\'c}}}, \bibinfo {author} {\bibfnamefont {S.}~\bibnamefont {Loire}},
  \bibinfo {author} {\bibfnamefont {V.~A.}\ \bibnamefont {Fonoberov}}, \ and\
  \bibinfo {author} {\bibfnamefont {P.}~\bibnamefont {Hogan}},\ }\bibfield
  {title} {\enquote {\bibinfo {title} {{A New Mixing Diagnostic and Gulf Oil
  Spill Movement}},}\ }\href {\doibase 10.1126/science.1194607} {\bibfield
  {journal} {\bibinfo  {journal} {Science}\ }\textbf {\bibinfo {volume}
  {330}},\ \bibinfo {pages} {486--489} (\bibinfo {year} {2010})}\BibitemShut
  {NoStop}%
\bibitem [{\citenamefont {Lolla}\ \emph {et~al.}(2014)\citenamefont {Lolla},
  \citenamefont {Lermusiaux}, \citenamefont {Ueckermann},\ and\ \citenamefont
  {Haley~Jr}}]{Lolla14}%
  \BibitemOpen
  \bibfield  {author} {\bibinfo {author} {\bibfnamefont {T.}~\bibnamefont
  {Lolla}}, \bibinfo {author} {\bibfnamefont {P.~F.~J.}\ \bibnamefont
  {Lermusiaux}}, \bibinfo {author} {\bibfnamefont {M.~P.}\ \bibnamefont
  {Ueckermann}}, \ and\ \bibinfo {author} {\bibfnamefont {P.~J.}\ \bibnamefont
  {Haley~Jr}},\ }\bibfield  {title} {\enquote {\bibinfo {title} {{Time-optimal
  path planning in dynamic flows using level set equations: theory and
  schemes}},}\ }\href {\doibase 10.1007/s10236-014-0757-y} {\bibfield
  {journal} {\bibinfo  {journal} {Ocean Dynamics}\ }\textbf {\bibinfo {volume}
  {64}},\ \bibinfo {pages} {1373--1397} (\bibinfo {year} {2014})}\BibitemShut
  {NoStop}%
\bibitem [{\citenamefont {Vese}\ and\ \citenamefont {Chan}(2002)}]{Vese02}%
  \BibitemOpen
  \bibfield  {author} {\bibinfo {author} {\bibfnamefont {L.~A.}\ \bibnamefont
  {Vese}}\ and\ \bibinfo {author} {\bibfnamefont {T.~F.}\ \bibnamefont
  {Chan}},\ }\bibfield  {title} {\enquote {\bibinfo {title} {{A multiphase
  level set framework for image segmentation using the Mumford and Shah
  model}},}\ }\href {\doibase 10.1023/A:1020874308076} {\bibfield  {journal}
  {\bibinfo  {journal} {International journal of computer vision}\ }\textbf
  {\bibinfo {volume} {50}},\ \bibinfo {pages} {271--293} (\bibinfo {year}
  {2002})}\BibitemShut {NoStop}%
\bibitem [{\citenamefont {Zhao}, \citenamefont {Osher},\ and\ \citenamefont
  {Fedkiw}(2001)}]{Zhao01}%
  \BibitemOpen
  \bibfield  {author} {\bibinfo {author} {\bibfnamefont {H.~K.}\ \bibnamefont
  {Zhao}}, \bibinfo {author} {\bibfnamefont {S.}~\bibnamefont {Osher}}, \ and\
  \bibinfo {author} {\bibfnamefont {R.}~\bibnamefont {Fedkiw}},\ }\bibfield
  {title} {\enquote {\bibinfo {title} {{Fast surface reconstruction using the
  level set method}},}\ }in\ \href {\doibase 10.1109/VLSM.2001.938900} {\emph
  {\bibinfo {booktitle} {{Variational and Level Set Methods in Computer Vision,
  2001. Proceedings. IEEE Workshop on}}}}\ (\bibinfo {year} {2001})\ pp.\
  \bibinfo {pages} {194--201}\BibitemShut {NoStop}%
\bibitem [{\citenamefont {Rudin}, \citenamefont {Osher},\ and\ \citenamefont
  {Fatemi}(1992)}]{Rudin92}%
  \BibitemOpen
  \bibfield  {author} {\bibinfo {author} {\bibfnamefont {L.~I.}\ \bibnamefont
  {Rudin}}, \bibinfo {author} {\bibfnamefont {S.}~\bibnamefont {Osher}}, \ and\
  \bibinfo {author} {\bibfnamefont {E.}~\bibnamefont {Fatemi}},\ }\bibfield
  {title} {\enquote {\bibinfo {title} {{Nonlinear total variation based noise
  removal algorithms}},}\ }\href {\doibase
  http://dx.doi.org/10.1016/0167-2789(92)90242-F} {\bibfield  {journal}
  {\bibinfo  {journal} {Physica D: Nonlinear Phenomena}\ }\textbf {\bibinfo
  {volume} {60}},\ \bibinfo {pages} {259--268} (\bibinfo {year}
  {1992})}\BibitemShut {NoStop}%
\bibitem [{\citenamefont {Sussman}, \citenamefont {Smereka},\ and\
  \citenamefont {Osher}(1994)}]{Sussman94}%
  \BibitemOpen
  \bibfield  {author} {\bibinfo {author} {\bibfnamefont {M.}~\bibnamefont
  {Sussman}}, \bibinfo {author} {\bibfnamefont {P.}~\bibnamefont {Smereka}}, \
  and\ \bibinfo {author} {\bibfnamefont {S.}~\bibnamefont {Osher}},\ }\bibfield
   {title} {\enquote {\bibinfo {title} {{A level set approach for computing
  solutions to incompressible two-phase flow}},}\ }\href {\doibase
  http://dx.doi.org/10.1006/jcph.1994.1155} {\bibfield  {journal} {\bibinfo
  {journal} {Journal of Computational physics}\ }\textbf {\bibinfo {volume}
  {114}},\ \bibinfo {pages} {146--159} (\bibinfo {year} {1994})}\BibitemShut
  {NoStop}%
\bibitem [{\citenamefont {Sethian}\ and\ \citenamefont
  {Smereka}(2003)}]{Sethian03}%
  \BibitemOpen
  \bibfield  {author} {\bibinfo {author} {\bibfnamefont {J.~A.}\ \bibnamefont
  {Sethian}}\ and\ \bibinfo {author} {\bibfnamefont {P.}~\bibnamefont
  {Smereka}},\ }\bibfield  {title} {\enquote {\bibinfo {title} {{Level set
  methods for fluid interfaces}},}\ }\href {\doibase
  10.1146/annurev.fluid.35.101101.161105} {\bibfield  {journal} {\bibinfo
  {journal} {Annual Review of Fluid Mechanics}\ }\textbf {\bibinfo {volume}
  {35}},\ \bibinfo {pages} {341--372} (\bibinfo {year} {2003})}\BibitemShut
  {NoStop}%
\bibitem [{\citenamefont {Leung}(2011)}]{Leung11}%
  \BibitemOpen
  \bibfield  {author} {\bibinfo {author} {\bibfnamefont {S.}~\bibnamefont
  {Leung}},\ }\bibfield  {title} {\enquote {\bibinfo {title} {{An Eulerian
  approach for computing the finite time Lyapunov exponent}},}\ }\href
  {\doibase http://dx.doi.org/10.1016/j.jcp.2011.01.046} {\bibfield  {journal}
  {\bibinfo  {journal} {Journal of computational physics}\ }\textbf {\bibinfo
  {volume} {230}},\ \bibinfo {pages} {3500--3524} (\bibinfo {year}
  {2011})}\BibitemShut {NoStop}%
\bibitem [{\citenamefont {You}\ and\ \citenamefont {Leung}(2015)}]{You15}%
  \BibitemOpen
  \bibfield  {author} {\bibinfo {author} {\bibfnamefont {G.}~\bibnamefont
  {You}}\ and\ \bibinfo {author} {\bibfnamefont {S.}~\bibnamefont {Leung}},\
  }\bibfield  {title} {\enquote {\bibinfo {title} {{A Fast Semi-Implicit Level
  Set Method for Curvature Dependent Flows with an Application to Limit Cycles
  Extraction in Dynamical Systems}},}\ }\href {\doibase
  10.4208/cicp.290414.231214a} {\bibfield  {journal} {\bibinfo  {journal}
  {Communications in Computational Physics}\ }\textbf {\bibinfo {volume}
  {18}},\ \bibinfo {pages} {203--229} (\bibinfo {year} {2015})}\BibitemShut
  {NoStop}%
\bibitem [{\citenamefont {You}\ and\ \citenamefont {Leung}(2014)}]{You14}%
  \BibitemOpen
  \bibfield  {author} {\bibinfo {author} {\bibfnamefont {G.}~\bibnamefont
  {You}}\ and\ \bibinfo {author} {\bibfnamefont {S.}~\bibnamefont {Leung}},\
  }\bibfield  {title} {\enquote {\bibinfo {title} {{An Eulerian method for
  computing the coherent ergodic partition of continuous dynamical systems}},}\
  }\href {\doibase http://dx.doi.org/10.1016/j.jcp.2014.01.034} {\bibfield
  {journal} {\bibinfo  {journal} {Journal of Computational Physics}\ }\textbf
  {\bibinfo {volume} {264}},\ \bibinfo {pages} {112--132} (\bibinfo {year}
  {2014})}\BibitemShut {NoStop}%
\bibitem [{\citenamefont {Fedkiw}, \citenamefont {Aslam},\ and\ \citenamefont
  {Xu}(1999)}]{Fedkiw99}%
  \BibitemOpen
  \bibfield  {author} {\bibinfo {author} {\bibfnamefont {R.~P.}\ \bibnamefont
  {Fedkiw}}, \bibinfo {author} {\bibfnamefont {T.}~\bibnamefont {Aslam}}, \
  and\ \bibinfo {author} {\bibfnamefont {S.}~\bibnamefont {Xu}},\ }\bibfield
  {title} {\enquote {\bibinfo {title} {{The ghost fluid method for deflagration
  and detonation discontinuities}},}\ }\href {\doibase
  http://dx.doi.org/10.1006/jcph.1999.6320} {\bibfield  {journal} {\bibinfo
  {journal} {Journal of Computational Physics}\ }\textbf {\bibinfo {volume}
  {154}},\ \bibinfo {pages} {393--427} (\bibinfo {year} {1999})}\BibitemShut
  {NoStop}%
\bibitem [{\citenamefont {Tsai}\ and\ \citenamefont {Osher}(2003)}]{Tsai03}%
  \BibitemOpen
  \bibfield  {author} {\bibinfo {author} {\bibfnamefont {R.}~\bibnamefont
  {Tsai}}\ and\ \bibinfo {author} {\bibfnamefont {S.}~\bibnamefont {Osher}},\
  }\bibfield  {title} {\enquote {\bibinfo {title} {{Review article: Level set
  methods and their applications in image science}},}\ }\href
  {http://projecteuclid.org/euclid.cms/1119655349} {\bibfield  {journal}
  {\bibinfo  {journal} {Commun. Math. Sci.}\ }\textbf {\bibinfo {volume} {1}},\
  \bibinfo {pages} {1--20} (\bibinfo {year} {2003})}\BibitemShut {NoStop}%
\bibitem [{\citenamefont {Karrasch}, \citenamefont {Huhn},\ and\ \citenamefont
  {Haller}(2014)}]{Karrasch15}%
  \BibitemOpen
  \bibfield  {author} {\bibinfo {author} {\bibfnamefont {D.}~\bibnamefont
  {Karrasch}}, \bibinfo {author} {\bibfnamefont {F.}~\bibnamefont {Huhn}}, \
  and\ \bibinfo {author} {\bibfnamefont {G.}~\bibnamefont {Haller}},\
  }\bibfield  {title} {\enquote {\bibinfo {title} {{Automated detection of
  coherent Lagrangian vortices in two-dimensional unsteady flows}},}\ }\href
  {\doibase 10.1098/rspa.2014.0639} {\bibfield  {journal} {\bibinfo  {journal}
  {Proceedings of the Royal Society of London A: Mathematical, Physical and
  Engineering Sciences}\ }\textbf {\bibinfo {volume} {471}},\ \bibinfo {pages}
  {20140639} (\bibinfo {year} {2014})}\BibitemShut {NoStop}%
\bibitem [{\citenamefont {Osher}\ and\ \citenamefont
  {Sethian}(1988)}]{Osher88}%
  \BibitemOpen
  \bibfield  {author} {\bibinfo {author} {\bibfnamefont {S.}~\bibnamefont
  {Osher}}\ and\ \bibinfo {author} {\bibfnamefont {J.~A.}\ \bibnamefont
  {Sethian}},\ }\bibfield  {title} {\enquote {\bibinfo {title} {{Fronts
  propagating with curvature-dependent speed: algorithms based on
  Hamilton-Jacobi formulations}},}\ }\href {\doibase
  http://dx.doi.org/10.1016/0021-9991(88)90002-2} {\bibfield  {journal}
  {\bibinfo  {journal} {Journal of computational physics}\ }\textbf {\bibinfo
  {volume} {79}},\ \bibinfo {pages} {12--49} (\bibinfo {year}
  {1988})}\BibitemShut {NoStop}%
\bibitem [{\citenamefont {Osher}\ and\ \citenamefont {Fedkiw}(2006)}]{Osher06}%
  \BibitemOpen
  \bibfield  {author} {\bibinfo {author} {\bibfnamefont {S.}~\bibnamefont
  {Osher}}\ and\ \bibinfo {author} {\bibfnamefont {R.}~\bibnamefont {Fedkiw}},\
  }\href@noop {} {\emph {\bibinfo {title} {{Level set methods and dynamic
  implicit surfaces}}}},\ Vol.\ \bibinfo {volume} {153}\ (\bibinfo  {publisher}
  {Springer Science \& Business Media},\ \bibinfo {year} {2006})\BibitemShut
  {NoStop}%
\bibitem [{\citenamefont {Evans}(2010)}]{Evans10}%
  \BibitemOpen
  \bibfield  {author} {\bibinfo {author} {\bibfnamefont {L.~C.}\ \bibnamefont
  {Evans}},\ }\href {https://books.google.ch/books?id=Xnu0o\_EJrCQC} {\emph
  {\bibinfo {title} {{Partial Differential Equations}}}},\ Graduate studies in
  mathematics\ (\bibinfo  {publisher} {American Mathematical Society},\
  \bibinfo {year} {2010})\BibitemShut {NoStop}%
\bibitem [{\citenamefont {Chang}\ \emph {et~al.}(1996)\citenamefont {Chang},
  \citenamefont {Hou}, \citenamefont {Merriman},\ and\ \citenamefont
  {Osher}}]{Chang96}%
  \BibitemOpen
  \bibfield  {author} {\bibinfo {author} {\bibfnamefont {Y.~C.}\ \bibnamefont
  {Chang}}, \bibinfo {author} {\bibfnamefont {T.~Y.}\ \bibnamefont {Hou}},
  \bibinfo {author} {\bibfnamefont {B.}~\bibnamefont {Merriman}}, \ and\
  \bibinfo {author} {\bibfnamefont {S.}~\bibnamefont {Osher}},\ }\bibfield
  {title} {\enquote {\bibinfo {title} {{A level set formulation of Eulerian
  interface capturing methods for incompressible fluid flows}},}\ }\href
  {\doibase http://dx.doi.org/10.1006/jcph.1996.0072} {\bibfield  {journal}
  {\bibinfo  {journal} {Journal of computational Physics}\ }\textbf {\bibinfo
  {volume} {124}},\ \bibinfo {pages} {449--464} (\bibinfo {year}
  {1996})}\BibitemShut {NoStop}%
\bibitem [{\citenamefont {Truesdell}\ and\ \citenamefont
  {Noll}(2004)}]{Truesdell04}%
  \BibitemOpen
  \bibfield  {author} {\bibinfo {author} {\bibfnamefont {C.}~\bibnamefont
  {Truesdell}}\ and\ \bibinfo {author} {\bibfnamefont {W.}~\bibnamefont
  {Noll}},\ }\href {\doibase 10.1007/978-3-662-10388-3_1} {\emph {\bibinfo
  {title} {{The Non-Linear Field Theories of Mechanics}}}},\ edited by\
  \bibinfo {editor} {\bibfnamefont {S.~S.}\ \bibnamefont {Antman}}\ (\bibinfo
  {publisher} {Springer Berlin Heidelberg},\ \bibinfo {address} {Berlin,
  Heidelberg},\ \bibinfo {year} {2004})\ Chap.\ \bibinfo {chapter} {The
  Non-Linear Field Theories of Mechanics}, pp.\ \bibinfo {pages}
  {1--579}\BibitemShut {NoStop}%
\bibitem [{\citenamefont {Csisz{\'a}r}\ and\ \citenamefont
  {Tusnady}(1984)}]{Csiszar84}%
  \BibitemOpen
  \bibfield  {author} {\bibinfo {author} {\bibfnamefont {I.}~\bibnamefont
  {Csisz{\'a}r}}\ and\ \bibinfo {author} {\bibfnamefont {G.}~\bibnamefont
  {Tusnady}},\ }\bibfield  {title} {\enquote {\bibinfo {title} {{Information
  geometry and alternating minimization procedures}},}\ }\href@noop {}
  {\bibfield  {journal} {\bibinfo  {journal} {Statistics and Decisions}\
  }\textbf {\bibinfo {volume} {Supplementary Issue 1}},\ \bibinfo {pages}
  {205.237} (\bibinfo {year} {1984})}\BibitemShut {NoStop}%
\bibitem [{\citenamefont {Chung}(1997)}]{Chung97}%
  \BibitemOpen
  \bibfield  {author} {\bibinfo {author} {\bibfnamefont {F.}~\bibnamefont
  {Chung}},\ }\href {https://books.google.ch/books?id=YUc38\_MCuhAC} {\emph
  {\bibinfo {title} {{Spectral graph theory}}}},\ Vol.~\bibinfo {volume} {92}\
  (\bibinfo  {publisher} {American Mathematical Soc.},\ \bibinfo {year}
  {1997})\BibitemShut {NoStop}%
\bibitem [{\citenamefont {Lankton}\ and\ \citenamefont
  {Tannenbaum}(2008)}]{Lankton08}%
  \BibitemOpen
  \bibfield  {author} {\bibinfo {author} {\bibfnamefont {S.}~\bibnamefont
  {Lankton}}\ and\ \bibinfo {author} {\bibfnamefont {A.}~\bibnamefont
  {Tannenbaum}},\ }\bibfield  {title} {\enquote {\bibinfo {title} {Localizing
  region-based active contours},}\ }\href {\doibase 10.1109/TIP.2008.2004611}
  {\bibfield  {journal} {\bibinfo  {journal} {IEEE Transactions on Image
  Processing}\ }\textbf {\bibinfo {volume} {17}},\ \bibinfo {pages}
  {2029--2039} (\bibinfo {year} {2008})}\BibitemShut {NoStop}%
\bibitem [{\citenamefont {Shi}\ and\ \citenamefont {Malik}(2000)}]{Shi00}%
  \BibitemOpen
  \bibfield  {author} {\bibinfo {author} {\bibfnamefont {J.~B.}\ \bibnamefont
  {Shi}}\ and\ \bibinfo {author} {\bibfnamefont {J.}~\bibnamefont {Malik}},\
  }\bibfield  {title} {\enquote {\bibinfo {title} {{Normalized cuts and image
  segmentation}},}\ }\href {<Go to ISI>://WOS:000089321500013} {\bibfield
  {journal} {\bibinfo  {journal} {IEEE Transactions on Pattern Analysis and
  Machine Intelligence}\ }\textbf {\bibinfo {volume} {22}},\ \bibinfo {pages}
  {888--905} (\bibinfo {year} {2000})}\BibitemShut {NoStop}%
\bibitem [{\citenamefont {Fu}\ \emph {et~al.}(2010)\citenamefont {Fu},
  \citenamefont {Chelton}, \citenamefont {Le~Traon},\ and\ \citenamefont
  {Morrow}}]{Fu10}%
  \BibitemOpen
  \bibfield  {author} {\bibinfo {author} {\bibfnamefont {L.}~\bibnamefont
  {Fu}}, \bibinfo {author} {\bibfnamefont {D.~B.}\ \bibnamefont {Chelton}},
  \bibinfo {author} {\bibfnamefont {P.-Y.}\ \bibnamefont {Le~Traon}}, \ and\
  \bibinfo {author} {\bibfnamefont {R.}~\bibnamefont {Morrow}},\ }\bibfield
  {title} {\enquote {\bibinfo {title} {{Eddy dynamics from satellite
  altimetry}},}\ }\href {http://dx.doi.org/10.5670/oceanog.2010.02} {\bibfield
  {journal} {\bibinfo  {journal} {Oceanography}\ }\textbf {\bibinfo {volume}
  {23}},\ \bibinfo {pages} {14--25} (\bibinfo {year} {2010})}\BibitemShut
  {NoStop}%
\bibitem [{\citenamefont {Sussman}\ \emph {et~al.}(1999)\citenamefont
  {Sussman}, \citenamefont {Almgren}, \citenamefont {Bell}, \citenamefont
  {Colella}, \citenamefont {Howell},\ and\ \citenamefont
  {Welcome}}]{Sussman99}%
  \BibitemOpen
  \bibfield  {author} {\bibinfo {author} {\bibfnamefont {M.}~\bibnamefont
  {Sussman}}, \bibinfo {author} {\bibfnamefont {A.~S.}\ \bibnamefont
  {Almgren}}, \bibinfo {author} {\bibfnamefont {J.~B.}\ \bibnamefont {Bell}},
  \bibinfo {author} {\bibfnamefont {P.}~\bibnamefont {Colella}}, \bibinfo
  {author} {\bibfnamefont {L.~H.}\ \bibnamefont {Howell}}, \ and\ \bibinfo
  {author} {\bibfnamefont {M.~L.}\ \bibnamefont {Welcome}},\ }\bibfield
  {title} {\enquote {\bibinfo {title} {An adaptive level set approach for
  incompressible two-phase flows},}\ }\href {\doibase
  http://dx.doi.org/10.1006/jcph.1998.6106} {\bibfield  {journal} {\bibinfo
  {journal} {Journal of Computational Physics}\ }\textbf {\bibinfo {volume}
  {148}},\ \bibinfo {pages} {81 -- 124} (\bibinfo {year} {1999})}\BibitemShut
  {NoStop}%
\bibitem [{\citenamefont {Persson}\ and\ \citenamefont
  {Strang}(2004)}]{Persson04}%
  \BibitemOpen
  \bibfield  {author} {\bibinfo {author} {\bibfnamefont {P.~O.}\ \bibnamefont
  {Persson}}\ and\ \bibinfo {author} {\bibfnamefont {G.}~\bibnamefont
  {Strang}},\ }\bibfield  {title} {\enquote {\bibinfo {title} {{A Simple Mesh
  Generator in MATLAB}},}\ }\href {\doibase 10.1137/S0036144503429121}
  {\bibfield  {journal} {\bibinfo  {journal} {SIAM Review}\ }\textbf {\bibinfo
  {volume} {46}},\ \bibinfo {pages} {329--345} (\bibinfo {year} {2004})},\
  \Eprint {http://arxiv.org/abs/http://dx.doi.org/10.1137/S0036144503429121}
  {http://dx.doi.org/10.1137/S0036144503429121} \BibitemShut {NoStop}%
\bibitem [{\citenamefont {Asay-Davis}\ \emph {et~al.}(2009)\citenamefont
  {Asay-Davis}, \citenamefont {Marcus}, \citenamefont {Wong},\ and\
  \citenamefont {de~Pater}}]{Asay09}%
  \BibitemOpen
  \bibfield  {author} {\bibinfo {author} {\bibfnamefont {X.~S.}\ \bibnamefont
  {Asay-Davis}}, \bibinfo {author} {\bibfnamefont {P.~S.}\ \bibnamefont
  {Marcus}}, \bibinfo {author} {\bibfnamefont {M.~H.}\ \bibnamefont {Wong}}, \
  and\ \bibinfo {author} {\bibfnamefont {I.}~\bibnamefont {de~Pater}},\
  }\bibfield  {title} {\enquote {\bibinfo {title} {{Jupiter's shrinking Great
  Red Spot and steady Oval BA: Velocity measurements with the Advection
  Corrected Correlation Image Velocimetry automated cloud-tracking method}},}\
  }\href {\doibase http://dx.doi.org/10.1016/j.icarus.2009.05.001} {\bibfield
  {journal} {\bibinfo  {journal} {Icarus}\ }\textbf {\bibinfo {volume} {203}},\
  \bibinfo {pages} {164--188} (\bibinfo {year} {2009})}\BibitemShut {NoStop}%
\bibitem [{url()}]{url_1}%
  \BibitemOpen
  \href@noop {} {\enquote {\bibinfo {title} {{Jupiter Global Map}},}\ }\bibinfo
  {howpublished} {\url{http://photojournal.jpl.nasa.gov}}\BibitemShut {NoStop}%
\bibitem [{\citenamefont {Sethian}(1996)}]{Sethian96}%
  \BibitemOpen
  \bibfield  {author} {\bibinfo {author} {\bibfnamefont {J.~A.}\ \bibnamefont
  {Sethian}},\ }\bibfield  {title} {\enquote {\bibinfo {title} {{Theory,
  algorithms, and applications of level set methods for propagating
  interfaces}},}\ }\href {\doibase 10.1017/S0962492900002671} {\bibfield
  {journal} {\bibinfo  {journal} {Acta Numerica}\ }\textbf {\bibinfo {volume}
  {5}},\ \bibinfo {pages} {309--395} (\bibinfo {year} {1996})}\BibitemShut
  {NoStop}%
\bibitem [{\citenamefont {Li}\ \emph {et~al.}(2005)\citenamefont {Li},
  \citenamefont {Xu}, \citenamefont {Gui},\ and\ \citenamefont {Fox}}]{Li05}%
  \BibitemOpen
  \bibfield  {author} {\bibinfo {author} {\bibfnamefont {C.}~\bibnamefont
  {Li}}, \bibinfo {author} {\bibfnamefont {C.}~\bibnamefont {Xu}}, \bibinfo
  {author} {\bibfnamefont {C.}~\bibnamefont {Gui}}, \ and\ \bibinfo {author}
  {\bibfnamefont {M.~D.}\ \bibnamefont {Fox}},\ }\bibfield  {title} {\enquote
  {\bibinfo {title} {{Level set evolution without re-initialization: a new
  variational formulation}},}\ }in\ \href {\doibase 10.1109/CVPR.2005.213}
  {\emph {\bibinfo {booktitle} {2005 IEEE Computer Society Conference on
  Computer Vision and Pattern Recognition (CVPR'05)}}},\ Vol.~\bibinfo {volume}
  {1}\ (\bibinfo {year} {2005})\ pp.\ \bibinfo {pages} {430--436 vol.
  1}\BibitemShut {NoStop}%
\bibitem [{\citenamefont {Sethian}(2001)}]{Sethian01}%
  \BibitemOpen
  \bibfield  {author} {\bibinfo {author} {\bibfnamefont {J.~A.}\ \bibnamefont
  {Sethian}},\ }\bibfield  {title} {\enquote {\bibinfo {title} {{Evolution,
  implementation, and application of level set and fast marching methods for
  advancing fronts}},}\ }\href {\doibase
  http://dx.doi.org/10.1006/jcph.2000.6657} {\bibfield  {journal} {\bibinfo
  {journal} {Journal of Computational Physics}\ }\textbf {\bibinfo {volume}
  {169}},\ \bibinfo {pages} {503--555} (\bibinfo {year} {2001})}\BibitemShut
  {NoStop}%
\bibitem [{\citenamefont {Zhang}\ \emph {et~al.}(2013)\citenamefont {Zhang},
  \citenamefont {Zhang}, \citenamefont {Song},\ and\ \citenamefont
  {Zhang}}]{Zhang13}%
  \BibitemOpen
  \bibfield  {author} {\bibinfo {author} {\bibfnamefont {K.}~\bibnamefont
  {Zhang}}, \bibinfo {author} {\bibfnamefont {L.}~\bibnamefont {Zhang}},
  \bibinfo {author} {\bibfnamefont {H.}~\bibnamefont {Song}}, \ and\ \bibinfo
  {author} {\bibfnamefont {D.}~\bibnamefont {Zhang}},\ }\bibfield  {title}
  {\enquote {\bibinfo {title} {{Reinitialization-free level set evolution via
  reaction diffusion}},}\ }\href {\doibase 10.1109/TIP.2012.2214046} {\bibfield
   {journal} {\bibinfo  {journal} {IEEE Transactions on Image Processing}\
  }\textbf {\bibinfo {volume} {22}},\ \bibinfo {pages} {258--271} (\bibinfo
  {year} {2013})}\BibitemShut {NoStop}%
\bibitem [{\citenamefont {Osher}\ and\ \citenamefont {Shu}(1991)}]{Osher91}%
  \BibitemOpen
  \bibfield  {author} {\bibinfo {author} {\bibfnamefont {S.}~\bibnamefont
  {Osher}}\ and\ \bibinfo {author} {\bibfnamefont {C.~W.}\ \bibnamefont
  {Shu}},\ }\bibfield  {title} {\enquote {\bibinfo {title} {{High-order
  essentially nonoscillatory schemes for Hamilton-Jacobi equations}},}\ }\href
  {http://www.jstor.org/stable/2157779} {\bibfield  {journal} {\bibinfo
  {journal} {SIAM Journal on Numerical Analysis}\ }\textbf {\bibinfo {volume}
  {28}},\ \bibinfo {pages} {907--922} (\bibinfo {year} {1991})}\BibitemShut
  {NoStop}%
\bibitem [{\citenamefont {Jiang}\ and\ \citenamefont {Peng}(2000)}]{Jiang00}%
  \BibitemOpen
  \bibfield  {author} {\bibinfo {author} {\bibfnamefont {G.~S.}\ \bibnamefont
  {Jiang}}\ and\ \bibinfo {author} {\bibfnamefont {D.}~\bibnamefont {Peng}},\
  }\bibfield  {title} {\enquote {\bibinfo {title} {{Weighted ENO schemes for
  Hamilton--Jacobi equations}},}\ }\href@noop {} {\bibfield  {journal}
  {\bibinfo  {journal} {SIAM Journal on Scientific computing}\ }\textbf
  {\bibinfo {volume} {21}},\ \bibinfo {pages} {2126--2143} (\bibinfo {year}
  {2000})}\BibitemShut {NoStop}%
\bibitem [{\citenamefont {Adalsteinsson}\ and\ \citenamefont
  {Sethian}(1995)}]{Adalsteinsson95}%
  \BibitemOpen
  \bibfield  {author} {\bibinfo {author} {\bibfnamefont {D.}~\bibnamefont
  {Adalsteinsson}}\ and\ \bibinfo {author} {\bibfnamefont {J.~A.}\ \bibnamefont
  {Sethian}},\ }\bibfield  {title} {\enquote {\bibinfo {title} {{A fast level
  set method for propagating interfaces}},}\ }\href {\doibase
  http://dx.doi.org/10.1006/jcph.1995.1098} {\bibfield  {journal} {\bibinfo
  {journal} {Journal of computational physics}\ }\textbf {\bibinfo {volume}
  {118}},\ \bibinfo {pages} {269--277} (\bibinfo {year} {1995})}\BibitemShut
  {NoStop}%
\bibitem [{\citenamefont {Whitaker}(1998)}]{Whitaker98}%
  \BibitemOpen
  \bibfield  {author} {\bibinfo {author} {\bibfnamefont {R.~T.}\ \bibnamefont
  {Whitaker}},\ }\bibfield  {title} {\enquote {\bibinfo {title} {{A level-set
  approach to 3D reconstruction from range data}},}\ }\href {\doibase
  10.1023/A:1008036829907} {\bibfield  {journal} {\bibinfo  {journal}
  {International journal of computer vision}\ }\textbf {\bibinfo {volume}
  {29}},\ \bibinfo {pages} {203--231} (\bibinfo {year} {1998})}\BibitemShut
  {NoStop}%
\bibitem [{\citenamefont {Lankton}(2009)}]{Lankton09}%
  \BibitemOpen
  \bibfield  {author} {\bibinfo {author} {\bibfnamefont {S.}~\bibnamefont
  {Lankton}},\ }\emph {\bibinfo {title} {{Localized Statisticical Models in
  Computer Vision}}},\ \href@noop {} {Ph.D. thesis},\ \bibinfo  {school}
  {Georgia Institute of Technology} (\bibinfo {year} {2009})\BibitemShut
  {NoStop}%
\end{thebibliography}%
\end{document}